\documentclass{article}
\usepackage{epsfig}
\def\openone{\leavevmode\hbox{\small1\kern-3.3pt\normalsize1}}
\def\tr{\mbox{tr\,}}

\def\re{\mbox{Re\,}}
\def\im{\mbox{Im\,}}

\begin{document}

\begin{center}

\begin{flushright}{}
{\bf nlin.IS/0104022}
\end{flushright}
\bigskip

{\Large\bf Adiabatic Interaction of $N$ Ultrashort Solitons:

\medskip
Universality of the Complex Toda Chain Model}

\bigskip

{\bf V. S. Gerdjikov}\\
{\sl Institute for Nuclear Research and Nuclear Energy, 1784
Sofia, Bulgaria}

\bigskip

{\bf E. V. Doktorov}\\ {\sl B. I. Stepanov Institute of Physics,
220072 Minsk, Belarus}

\bigskip

{\bf J. Yang}\\ {\sl Department of Mathematics and Statistics,
University of Vermont, \\ Burlington, VT 05401,  USA}

\begin{abstract}
Using the Karpman-Solov'ev method we derive the equations for the
$2 $-soliton adiabatic interaction for solitons of the modified
nonlinear Schr\"odin\-ger  equation (MNSE).  Then we generalize
these equations to the case of $N$ interacting solitons with
almost equal velocities and widths. On the basis of this result
we prove that the $N $ MNSE-soliton train interaction ($N>2$) can
be modeled by the completely integrable complex Toda chain (CTC).
This is an argument in favor of universality of the complex Toda
chain which was previously shown to model the soliton train
interaction for nonlinear Schr\"odinger solitons. The
integrability of the CTC is used to describe all possible
dynamical regimes of the $N $-soliton trains which include
asymptotically free propagation of all $N $ solitons, $N
$-soliton bound states, various mixed regimes, etc. It allows
also to describe analytically the manifolds in the $4N
$-dimensional space of initial soliton parameters which are
responsible for each of the regimes mentioned above. We compare
the results of the CTC model with the numerical solutions of the
MNSE  for $2 $ and $3 $-soliton interactions and find a very good
agreement.

\end{abstract}
\end{center}

\section{Introduction}\label{sec:In}

The analytical description of the dynamics of picosecond solitons
in single-mode nonlinear fibers is based on the nonlinear
Schr\"odinger equation (NSE)~\cite{Agraw,Kod1}. The NSE serves as
a very good integrable model admitting comprehensive investigation
in the framework of the inverse spectral transform
(IST)~\cite{NMPZ}. IST provides the complete analytical
description of the soliton interaction in a generic case of
asymptotically free $N$ solitons moving with pair-wise different
velocities~\cite{NMPZ,Takh-Fad}. On the other hand, the
practically important case, especially in a soliton-based fiber
transmission, deals with the so-called $N$-soliton trains, i.e.,
with an ordered sequence of $N$ ($N\ge 2$) solitons which are
spaced apart almost equally and have almost (or exactly) equal
amplitudes and velocities. In a number of recent
papers~\cite{Gerd1,Gerd2,Gerd3,Gerd4}, an effective formalism was
developed for studying the dynamics of well-separated NSE
solitons within the $N$-soliton train. This approach is based on
a generalization of the two-soliton quasiparticle method by
Karpman and Solov'ev~\cite{KS} to the case of $N$ solitons. In the
framework of this approach, the soliton interaction is governed
by a dynamical system for $4N$ soliton parameters. Such an
approximation is called adiabatic because interaction between the
solitons is displayed as a slow deformation of their parameters,
a possible presence of radiation being ignored. It is important
to realize that the above generalization from two to $N$ solitons
is nontrivial because of lack of the superposition principle for
the nonlinear dynamical system.

Under some additional restrictions imposed on the soliton parameters,
which ensure the validity of the adiabatic approximation, the above
dynamical system is reduced to the complex Toda chain (CTC) equations with
$N$ nodes~\cite{Arn}.  Extensive use of the fact that the CTC is a
completely integrable model permits to classify soliton parameter regions
with different asymptotic regimes of the $N$-soliton
train~\cite{Gerd1,Gerd2,Gerd3,Gerd4}. It was also shown in~\cite{Gerd4}
that the CTC can be associated with any equation from the NSE hierarchy.

One of the purposes in the optical fiber soliton communication is
to achieve as high of a bit rate as possible. A natural way in
this direction is the use of shorter optical pulses. It should be
noted, however, that when dealing with ultrashort optical pulses
with duration $\le 100$ fs, the NSE should be modified to take
into account some additional effects, such as the nonlinearity
dispersion, the intrapulse Raman scattering and the higher-order
dispersion~\cite{Agraw}. As a rule, the extra terms added to the
NSE violate its integrability. On the other hand, if these
additional terms are small, the IST-based soliton perturbation
theory is usualy treated as the relevant method to account for their
influence on the soliton behavior~\cite{Kaup,KM,Kiv-Mal}.

It is remarkable that adding a term accounting for the nonlinearity
dispersion to the NSE preserves the integrability of the equation. In
other words, the modified nonlinear Schr\"odinger equation (MNSE)
\begin{equation}\label{eq:1}
iu_t+{1\over 2}u_{xx}+i\alpha\left(|u|^2u\right)_x+|u|^2u=0
\end{equation}
is still integrable by means of IST, though the associated
spectral problem (the so called Wadati-Konno-Ichikawa spectral
problem~\cite{WKI}, or quadratic bundle) does not belong to the
familiar Zakharov-Shabat class. The parameter $\alpha$ in
(\ref{eq:1}) governs the strength of the nonlinearity dispersion.
The case $\alpha=0$ corresponds to NSE. Thereby, the effect of
the nonlinearity dispersion is considered non-perturbatively in
(\ref{eq:1}).  Moreover, we have to stress that it is the
completely integrable model (\ref{eq:1}) that should be
considered as a true starting point for analytical investigation
of subpicosecond soliton dynamics. Indeed, it was shown
in~\cite{Ohkuma} that numerical simulation of the soliton
propagation according to the MNSE (\ref{eq:1}) revealed various
kinds of dynamical behavior which cannot be accounted for by
treating the nonlinearity dispersion term of the MNSE
(\ref{eq:1}) as a perturbation term in the NSE. Analogous idea
in treating the perturbed NSE  was developed by Kodama
and Hasegawa in~\cite{Kod}. There the NSE with perturbations like the
third order dispersion, nonlinear gain and nonlinear dispersion
was treated as a perturbed higher order NSE.

The relevance of Eq. (\ref{eq:1}) to the problem of ultrashort
pulse propagation in fibers was demonstrated
in~\cite{Tzoar,Anders}. MNSE (\ref{eq:1}) is also used in plasma
physics~\cite{Mio} and is relevant for description of a deformed
continuous Heisenberg ferromagnet~\cite{India}. It is the Alfv\'en
waves in magnetized plasma where the first successful application
of IST to the quadratic bundle was achieved on an example of the
derivative NSE~\cite{DNLS}  which is Eq. (\ref{eq:1}) without the
last term. Both equations are interrelated by a gauge-like
transformation, see, for example,~\cite{Mihal,Gerd-Iv}. The
soliton solutions and the Hamiltonian structures of the MNSE were
obtained for the first time in~\cite{Gerd-Iv}. $N$-soliton
solutions were further rederived by different methods: by IST
using the above relation with the derivative NSE~\cite{Vysl}, by
B\"acklund and Darboux transformations~\cite{Rao}, by technique of
determinant calculations~\cite{Chen}, by the Hirota
method~\cite{Lin}, by the $\bar\partial$-method~\cite{I&Val}. It
should be noted that the solutions obtained in these papers refer
to the general case of asymptotically free solitons and being
exact were too complicated for practical use.

Recently, a novel parametrization for the MNSE solitons was
proposed within the framework of the Riemann-Hilbert formulation
of IST~\cite{Val&I}. The convenient parametrization of the MNSE
soliton facilitated the development of an effective adiabatic
soliton perturbation theory for the MNSE which is able to take into
account  non-zero terms in the right-hand side of (\ref{eq:1}),
see~\cite{Val&I}.

The next natural step is to derive dynamical equations of the
Karpman-Solov'ev type for the adiabatic evolution of the soliton
parameters for the MNSE $N$-soliton train. Several questions
arise in the process of solving this problem. Is it possible to
associate an $N$-node chain model, like the CTC, with this
dynamical system? Will this chain model be different from the CTC
and, therefore, is the CTC valid only for the NSE hierarchy? How
well do the numerical simulations of the MNSE with adiabatic
$N$-soliton train initial conditions agree with the chain-like
model predictions? All these questions will be answered below.

The purpose of this paper is to derive a dynamical system for the
$4N$ soliton parameters for the MNSE $N$-soliton train. To this
end we will generalize to the quadratic bundle the similar
investigations performed for the NSE. In the next section we
apply the Karpman-Solov'ev approach to the MNSE (\ref{eq:1}) and
derive the dynamical system for the 2-soliton train. In Sec. 3 we
show how this result can be generalized to the MNSE $N$-soliton
train for $N>2$. We  show in Sec. 4 that after some
additional assumptions the corresponding dynamical system for the
soliton parameters acquires the form of the CTC. Thus we find
that the CTC is characteristic not only for the NSE hierarchy
\cite{Gerd4}, but has wider field of applications.  This is an
argument in favor of its universality.

In Section 5 we show how the integrability of the CTC can be used
to determine the dynamical regimes of the $N $-soliton trains. We
demonstrate on the examples of $N=2 $ and $N=3 $ how one can
describe analytically the manifolds in the $4N $-dimensional
space of initial soliton parameters which are responsible for the:
i)~$N $-soliton bound state regime; ii)~asymptotically free
regimes; iii)~various mixed regimes, etc. Although the analysis
follows closely the ideas developed in \cite{Gerd1,Gerd2,Gerd3}
the description of the corresponding manifolds differs from the
ones for the NSE soliton trains. The reason for this lies in the
fact that the CTC fields $Q_j(t) $ are parametrized in a different
way; in particular, $\im Q_j $ depend not only on the soliton
phases $\delta _j $ (as is the case for the NSE) but also on the
soliton amplitudes.

In Secton 6 predictions of the CTC model are compared with the
numerical results from the MNSE  and find an excellent match for
most regimes with $N=2 $ and $N=3 $. We found some disagreement
between the CTC and numerical MNSE solution in the regimes when
CTC predicts a very slow soliton separation.

\section{$2 $-soliton interactions for the MNSE}\label{sec:II}

First of all we summarize the basic results concerning the soliton
solution of the MNSE (\ref{eq:1})~\cite{Val&I}. This equation
admits the Lax representation
\begin{eqnarray}
\Phi_x &=& -\frac{2i}{\alpha} \left( k^2-\frac{1}{4} \right) \left[
\sigma_3,\Phi \right] + 2ikQ\Phi,
\label{eq:2} \\
\Phi_t&=&-\frac{4i}{\alpha^2}\left(k^2-\frac{1}{4}\right)^2
\left[\sigma_3,\Phi\right] \nonumber\\
&+&\left(\frac{4i}{\alpha}k^3Q+2ik^2Q^2\sigma_3-\frac{i}{\alpha}
kQ+kQ_x\sigma_3-2i\alpha kQ^3\right)\Phi. \nonumber
\end{eqnarray}
Here the Hermitian matrix $Q=\pmatrix{0 &u\cr \bar u& 0\cr}$
stands for the potential of the spectral problem (\ref{eq:2}), $k$
is a spectral parameter. There exist various parametrizations of
the soliton solution of the MNSE, the first one having been
proposed in~\cite{Gerd-Iv}. We follow here the parametrization
given in~\cite{Val&I} which was proven  to be useful for practical
calculations and admits a simple (though nontrivial) reduction to
the NSE for $\alpha\to0$. Namely, the MNSE soliton solution is
written as follows:
\begin{equation}
u_s(z,t)=i\frac{\nu}{\alpha}\frac{ke^{-z}+\bar ke^z}{(ke^z+\bar
ke^{-z})^2}e^{i\phi}. \label{eq:3}
\end{equation}
Here $k=k_R-ik_I,\quad k_I>0, \quad \lambda\equiv
4k^2-1=\mu-i\nu$,
\begin{eqnarray}\label{eq:4}
z &=& -\frac{\nu}{\alpha}\left(x-\xi(t)\right), \qquad
\phi=\frac{\mu}{\nu}z+\delta(t), \nonumber\\
\xi(t)&=&-\frac{\mu}{\alpha}t+\xi_{(0)}, \qquad
\delta(t)=\frac{1}{2\alpha^2}(\mu^2+\nu^2)t+\delta_{(0)}.
\end{eqnarray}
It should be stressed that the soliton (\ref{eq:3}) is not of the
hyperbolic-secant type with a real argument, characteristic
for the NSE. It is specified by four real parameters $\mu$, $\nu$,
$\xi_{(0)}$, and $\delta_{(0)}$ with $(-\mu/\alpha)$ being the
soliton velocity, $(\alpha/\nu)$ is its width, $\delta_{(0)}$ and
$\xi_{(0)}$ are initial phase and position of the soliton. To
carry out the limit reduction to the NSE, one should decompose
the spectral parameter in the following manner:
\begin{equation}
k=\frac{1}{2}-\frac{\alpha}{2}\left(\mu^{\rm NSE}+i\nu^{\rm
NSE}\right)+{\cal O}(\alpha^2), \qquad \alpha\to 0, \label{eq:5}
\end{equation}
which gives $(-\mu/\alpha)\to 2\mu^{\rm NSE}$ and
$(\alpha/\nu)\to(2\nu^{\rm NSE})^{-1}$, as should be.

If there is a small perturbation in a system described by the MNSE, we
will deal with a perturbed MNSE:
\begin{equation}
iu_t+{1\over 2}u_{xx}+i\alpha\left(|u|^2u\right)_x+|u|^2u=r(x,t),
\label{eq:6}
\end{equation}
where $r(x,t)$ describes a functional form of the perturbation.
In what follows we will restrict ourselves to the adiabatic
approximation of the soliton perturbation theory. In other words,
we suppose that a perturbation causes a slow variation of the
soliton parameters only.  The evolution equations for the
perturbation-induced soliton parameters are given in~\cite{Val&I}.
Here we write them in terms of the parameters (\ref{eq:4}). The
key equation has a very simple form:
\begin{equation}
\frac{\mbox{d}k}{\mbox{d}t}=\frac{i}{2}\alpha k^2\int_{-\infty}^\infty
\frac{R_+e^z}{(ke^{-z}+\bar ke^z)^2}\mbox{d}z, \label{eq:7}
\end{equation}
where $R_\pm=\exp[-i\phi(z,t)]r(z,t)\pm\exp[i\phi(-z,t)]\bar
r(-z,t)$.  Taking into account (\ref{eq:4}), we obtain
\begin{eqnarray}
\frac{\mbox{d}\mu}{\mbox{d}t}&=&2i\alpha\int_{-\infty}^\infty
\frac{k^3e^z-\bar k^3e^{-z}}{(ke^{-z}+\bar ke^z)^2}R_+\mbox{d}z,
\label{eq:8}\\
\frac{\mbox{d}\nu}{\mbox{d}t}&=&-2\alpha\int_{-\infty}^\infty
\frac{k^3e^z+\bar k^3e^{-z}}{(ke^{-z}+\bar ke^z)^2}R_+\mbox{d}z.
\label{eq:9}
\end{eqnarray}
Evolution of $\xi$ and $\delta$ is given by the following formulae:
\begin{eqnarray}\label{eq:10}
\frac{\mbox{d}\xi}{\mbox{d}t}&=&-\frac{\mu}{\alpha}-\frac{1}{\alpha\nu}
\left(\int_0^t\mbox{d}t'\mu(t')\right)\frac{\mbox{d}\nu}{\mbox{d}t}
\nonumber\\
& +& \frac{4\alpha^2}{\nu^2}\int_{-\infty}^\infty\frac{
\mbox{d}zR_-}{(ke^{-z}+\bar ke^z)^2}\left[z\left(k^3e^z+\bar
k^3e^{-z}\right)+\frac{i\nu}{8}\left(ke^z+\bar ke^{-z}\right)\right], \\
\frac{\mbox{d}\delta}{\mbox{d}t} &=& \frac{\mu^2+\nu^2}{2\alpha^2}+
\frac{1}{\alpha^2\nu}\left(\int_0^t\mbox{d}t'\mu(t')\right)
\left(\mu\frac{\mbox{d}\nu}{\mbox{d}t}-\nu\frac{\mbox{d}\mu}
{\mbox{d}t}\right) \nonumber\\
&+& \frac{4i\alpha}{\nu} \int_{-\infty}^\infty\frac{\mbox{d}zR_-}
{(ke^{-z}+\bar ke^z)^2}\Big[|k^2|\left(ke^{-z}+\bar k e^z\right)
\nonumber\\
&-& \frac{1}{8}\left(\bar\lambda ke^z-\lambda\bar ke^{-z}\right)
+\frac{iz}{\nu}\left(\bar\lambda k^3e^z+\lambda\bar k^3e^{-z}\right)
\Big]. \label{eq:11}
\end{eqnarray}
It should be noted that for the symmetric perturbations obeying
the condition $\exp[-i\phi(z,t)]r(z,t)=\exp[i\phi(-z,t)]\bar
r(-z,t)$, i.e., $R_-=0$, the complicated integrals in the right
hand sides of (\ref{eq:10}) and (\ref{eq:11}) disappear.

Now we have all the necessary information
to derive the Karpman-Solov'ev-like
dynamical system of equations for the adiabatic interaction of two
well-separated MNSE solitons. Below we will formulate more precisely the
condition of sufficient separability of solitons.  We suppose that a
two-soliton solution to the MNSE (\ref{eq:1}) is well approximated by the
sum of two MNSE solitons:
\begin{equation}
u(x,t)=u_1(z_1,t)+u_2(z_2,t), \label{eq:12}
\end{equation}
where $u_j(z_j,t)$, $j=1,2$, is given by (\ref{eq:3}) with
\begin{eqnarray}\label{eq:nn}
z_j&=&-\frac{\nu_j}{\alpha}(x-\xi_j), \qquad \phi_j=\frac{\mu_j}{\nu_j}
z_j+\delta_j. \nonumber\\
\xi_j(t)&=&-\frac{1}{\alpha}\int_0^t{\rm d} t'\mu_j(t')+\xi_{j0},
\quad \delta_j(t)=\frac{1}{2\alpha^2}\int_0^t{\rm d}
t'\left(\mu_j^2(t')+\nu_j^2(t')\right)+\delta_{j0},\nonumber
\end{eqnarray}
where we took into account for the possible evolution of $\mu_j$
and $\nu_j$. Substituting now (\ref{eq:12}) into the MNSE
(\ref{eq:1}), it is easy to see that, because of the
nonlinearity, each soliton feels the presence of the other one
and the interaction is described by the perturbed MNSE
\begin{equation}
iu_{jt}+{1\over 2}u_{jxx}+i\alpha\left(|u_j|^2u_j\right)_x+|u_j|^2u_j=r_j,
\label{eq:13}
\end{equation}
where
\begin{equation}
r_j=-i\alpha\left(2|u_j|^2u_{3-j}+u_j^2\bar u_{3-j}\right)_x-
\left(2|u_j|^2u_{3-j}+u_j^2\bar u_{3-j}\right).\label{eq:14}
\end{equation}
It should be stressed that the perturbation (\ref{eq:14}) arises
effectively as a result of treating two-soliton solution as the sum
(\ref{eq:12}) of the one-soliton ones.

Now we formulate the conditions which provide the representation
(\ref{eq:12}) as a two-soliton solution of the MNSE (\ref{eq:1}). At first
we express $z_2$ in terms of $z_1$,
\[
z_2=\left(1+\frac{\nu_2-\nu_1}{\nu_1}\right)z_1+
\frac{\nu_2}{\alpha}(\xi_2-\xi_1).
\]
We suppose that solitons have almost equal widths, i.e.,
\begin{equation}
\frac{|\nu_2-\nu_1|}{\nu_0}\ll 1, \label{eq:15}
\end{equation}
where $\nu_0=(1/2)(\nu_1+\nu_2)$. Hence, we have
\begin{equation}
z_2-z_1\simeq\frac{\nu_0}{\alpha}(\xi_2-\xi_1). \label{eq:16}
\end{equation}
Calculation of the overlap integral $|\int_{-\infty}^\infty
u_1(z_1,t)u_2(z_2,t){\rm d} x|$ (or, equivalently,
$\int_{-\infty}^\infty{\rm d} x|u_1(z_1,t)u_2(z_2,t)|$) gives an
expression containing the factor
$\exp[-(\nu_0/\alpha)(\xi_2-\xi_1)]\equiv\epsilon$ for
$\xi_2>\xi_1$. Just this exponential factor determines a measure
of overlapping neighboring solitons. We take in the following
\begin{equation}
\frac{\nu_0}{\alpha}|\xi_2-\xi_1|\gg 1 \label{eq:17}
\end{equation}
(or $\epsilon \ll 1$) which means weak overlapping between the solitons.

Let us consider now the phase difference
$\phi_2-\phi_1=(\mu_2/\nu_2)z_2-(\mu_1/\nu_1)z_1+\delta_2-\delta_1$.
Accounting for (\ref{eq:15}) and (\ref{eq:16}) we may write
\[
\phi_2-\phi_1=\frac{1}{\nu_2}\left[\mu_2-\left(1+\frac{\nu_2-\nu_1}
{\nu_1}\right)\mu_1\right] z_1+ \frac{\mu_2}{\alpha}\frac{\nu_0}
{\nu_2}(\xi_2-\xi_1)+\delta_2-\delta_1.
\]
Since we consider solitons moving with small relative velocities we
assume:
\begin{equation}
\frac{|\mu_2-\mu_1|}{\nu_0}\ll 1. \label{eq:18}
\end{equation}
Then the phase difference will not contain the $z$-dependence.
Furthermore,
\[
\frac{\mu_2}{\alpha}\frac{\nu_0}{\nu_2}(\xi_2-\xi_1)=
\frac{\mu_2}{\alpha}\left(1+\frac{\nu_0-\nu_2}{\nu_2}\right)(\xi_2-\xi_1).
\]
As the last condition we suppose
\begin{equation}
|\nu_j-\nu_0|(\xi_2-\xi_1)\ll 1, \label{eq:19}
\end{equation}
hence, the phase difference takes the form
\[
\phi_2-\phi_1=\frac{\mu_0}{\alpha}(\xi_2-\xi_1)+\delta_2-\delta_1.
\]
Therefore, the conditions (\ref{eq:15}), (\ref{eq:17}), (\ref{eq:18}),
(\ref{eq:19}) provide a possibility to consider a two-soliton solution of
the MNSE (\ref{eq:1}) as the sum of the form (\ref{eq:12}).

To derive the Karpman-Solov'ev-like equations for the soliton parameters,
we use Eqs. (\ref{eq:7})--(\ref{eq:11}) with the perturbation
(\ref{eq:14}).  Accounting for the above conditions, we obtain after
simple but tedious calculations ($j=1,2$):
\begin{equation}
\frac{\mbox{d}\lambda_j}{\mbox{d}t}=(-1)^j 4
k_j\left(\frac{w_j\nu_j} {\alpha}\right)^2
\frac{w_{3-j}\nu_{3-j}}{\bar k_{3-j}}e^{-\Delta-i\psi},
\label{eq:20}
\end{equation}
\begin{equation}
\frac{\mbox{d}\mu_j}{\mbox{d}t}=(-1)^j\frac{2}{\alpha^2}\nu_j^2
\nu_{3-j}\left(w_j^2 w_{3-j}\frac{k_j}{\bar
k_{3-j}}e^{-i\psi}+\bar w_j^2\bar w_{3-j}\frac{\bar
k_j}{k_{3-j}}e^{i\psi}\right)e^{-\Delta}, \label{eq:21}
\end{equation}
\begin{equation}
\frac{\mbox{d}\nu_j}{\mbox{d}t}=(-1)^{j+1}\frac{2i}{\alpha^2}\nu_j^2
\nu_{3-j}\left(w_j^2 w_{3-j}\frac{k_j}{\bar
k_{3-j}}e^{-i\psi}-\bar w_j^2\bar w_{3-j}\frac{\bar
k_j}{k_{3-j}}e^{i\psi}\right)e^{-\Delta}, \label{eq:22}
\end{equation}
\begin{eqnarray}
\frac{\mbox{d}\xi_j}{\mbox{d}t}&=&-\frac{1}{\alpha}\mu_j + (-1)^j
\frac{2i}{\alpha^3}\nu_j\nu_{3-j} \left(\int_0^t\mbox{d}t'\mu_j(t')\right)
\nonumber\\
&\times & \left(w_j^2 w_{3-j}\frac{k_j}{\bar k_{3-j}}
e^{-i\psi}-\bar w_j^2\bar w_{3-j}\frac{\bar k_k}{k_{3-j}}
e^{i\psi}\right)e^{-\Delta} \nonumber \\
&+&\frac{i}{\alpha}\nu_{3-j}\Bigg(\left[\left(1+\bar w_j^2\right)
\left(1-2\bar w_j^2\right)+4is_j\right] w_j^2 w_{3-j}\frac{k_j}
{\bar k_{3-j}}e^{-i\psi} \nonumber \\
&-&\left[\left(1+w_j^2\right)\left(1-2w_j^2\right)-4is_j\right]
\bar w_j^2\bar w_{3-j}\frac{\bar k_j}{k_{3-j}}e^{i\psi}\Bigg)
e^{-\Delta},\label{eq:23}
\end{eqnarray}
\begin{eqnarray}
\frac{\mbox{d}\delta_j}{\mbox{d}t}&=&\frac{1}{2\alpha^2}
\left(\mu_j^2+\nu_j^2\right) - (-1)^j\frac{2i}{\alpha^4}
\nu_j\nu_{3-j}\left(\int_0^t\mbox{d} t'\mu_j(t') \right) \nonumber\\
&\times &\left(\lambda_j w_j^2 w_{3-j}\frac{k_j}{\bar
k_{3-j}}e^{-i\psi}-\bar\lambda_j\bar w_j^2\bar w_{3-j}
\frac{\bar k_j}{k_{3-j}}e^{i\psi}\right)e^{-\Delta} \nonumber \\
&-&\frac{i}{\alpha^2}\bar\lambda_j\nu_{3-j}\Bigg(\left[
\left(1+\bar w_j^2\right)\left(1-2\bar
w_j^2\right)+4is_j\right]w_j^2
w_{3-j}\frac{k_j}{\bar k_{3-j}}e^{-i\psi} \nonumber \\
&-&\left[\left(1+w_j^2\right)\left(1-2w_j^2\right)-4is_j\right]
\bar w_j^2\bar w_{3-j}\frac{\bar k_j}{k_{3-j}}e^{i\psi}\Bigg)
e^{-\Delta}.\label{eq:24}
\end{eqnarray}
Above we used the notations
\begin{eqnarray*}
w_j = {k_j\over \bar k_j} = \exp(-2is_j), &\qquad & \Delta= {\nu_0
\over \alpha} |\xi_2-\xi_1|, \\
\psi = {\mu_0 \over \alpha} (\xi_2-\xi_1)+\delta_2-\delta_1, & \qquad &
s_j={1\over 2}  \arctan {\nu _j \over 1+\mu _j} ,
\end{eqnarray*}
where the last relation follows from $\lambda _j=4k_j^2-1 =\mu _j -i\nu _j
$.

Equations (\ref{eq:21})--(\ref{eq:24}) are the analog of the
Karpman-Solov'ev equations in the case of the adiabatic
interaction of two well-separated MNSE solitons and reduce to the
NSE dynamical system in the limit (\ref{eq:5}).

The dynamical system  (\ref{eq:21})--(\ref{eq:24}) is rather
complicated and needs further simplification to perform its
analytical investigation.  Integrable approximation is of special
importance, and finding such an approximation is one of our
purposes. But first of all we will generalize these equations to
the case of $N$ MNSE solitons.

\section{$N $-soliton train interactions for the MNSE}\label{sec:3}

Since the Karpman-Solov'ev-like dynamical system is nonlinear, it
does not allow the superposition principle. It is physically
clear because in the case of the $N$-soliton train with $N\ge3$ a
middle soliton will be influenced by its neighbors from both
sides. Hence, it is not possible to describe the interaction of
$N\ge3$ solitons within the framework of two-soliton interaction
like (\ref{eq:21})--(\ref{eq:24}).

The first remark we should keep in mind is that the interaction
force between the solitons is of the order of their overlap.
Therefore, we can take into account only the nearest neighbor
interaction. Indeed, for the $N$-soliton train we assume that
\begin{equation}
u=\sum_{j=1}^Nu_j \label{eq:25}
\end{equation}
where $u_j$ is the MNSE soliton (\ref{eq:3}) whose center of mass
is located at $\xi_j$. Assume that $\xi_1<\xi_2<\dots<\xi_N$.
Inserting this {\it ansatz} into the cubic term of the MNSE
(\ref{eq:1}) gives
\begin{equation}
|u^2|u=\sum_{j=1}^N|u_j^2|u_j+\sum_{j\ne l}\left(|u_j^2|u_l+2u_j^2
\bar u_l\right)+\sum_{j\ne l\ne n}u_j\bar u_lu_n. \label{eq:26}
\end{equation}
Straightforward analysis shows that the integrals in
(\ref{eq:7})--(\ref{eq:11}) corresponding to each type of terms in
(\ref{eq:26}) are of the following order of magnitude:
\begin{eqnarray}
|u_k|^2 u_m , \quad u_k^2 \bar u_m \qquad &\leftrightarrow &
\qquad {\cal O}(\epsilon^{-|k-m|}), \nonumber\\
u_j \bar u_l u_n , \quad j<l<n \qquad &\leftrightarrow & \qquad
{\cal O}(\epsilon^{-|j-l|-|l-n|}).\nonumber
\end{eqnarray}
Here $\epsilon$ is of the order of
$\exp[-(\nu_0/\alpha)|\xi_j-\xi_l|]$ for $|j-l|=1$. Because we
keep only terms of the order of $\epsilon$, we see that it is
enough to take into account only terms with $|j-l|=1$. In other
words, the 'triple' terms like $u_j \bar u_l u_n$ can be
neglected. Quite analogous is the situation with the cubic terms
containing $x$-derivative.

Secondly, as in Sec. 2, we pose the conditions
\[
|\nu_j-\nu_l|\ll \nu_0, \quad |\mu_j-\mu_l|\ll \nu_0, \quad
\nu_0|\xi_{0j}-\xi_{0l}|\gg 1, \quad
|\nu_j-\nu_l||\xi_{0j}-\xi_{0l}|\ll 1,
\]
where $\nu_0=N^{-1}\sum_{j=1}^N\nu_j$, $\mu_0=N^{-1}\sum_{j=1}^N\mu_j$.
They mean that we consider the chain-like configuration of $N$ solitons
with equal or nearly equal velocities and widths. Substituting the soliton
solutions (\ref{eq:3}) into the perturbation
\begin{equation}
r_j=-\sum_{l=j\pm1}\left[i\alpha\left(2|u_j|^2u_l+u_j^2\bar
u_l\right)_x+\left(2|u_j|^2u_l+u_j^2\bar u_l\right)\right]
\label{eq:27}
\end{equation}
and calculating the integrals in (\ref{eq:7}), we obtain
\begin{equation}
\frac{{\rm d}\lambda_j}{{\rm
d}t}=-k_j\left(\frac{2w_j\nu_j}{\alpha} \right)^2\sum_{l=j\pm
1}s_{lj} \frac{w_l\nu_l}{\bar
k_l}e^{-|\Delta_{lj}|}e^{-is_{lj}\psi_{lj}}.\label{eq:28}
\end{equation}
where
\begin{eqnarray}\label{eq:28'}
s_{lj}&=& \left\{ \begin{array}{ll} 1 & \mbox{for} \quad l = j+1, \\
-1 & \mbox{for} \quad l = j-1, \end{array}  \right. \qquad
\Delta_{lj}=\frac{\nu_0}{\alpha} |\xi_l-\xi_j| , \nonumber\\
\psi_{lj}&\equiv & \psi _l  -\psi _j = \frac{\mu_0}{\alpha}
(\xi_l-\xi_j)+\delta_l-\delta_j.
\end{eqnarray}
The corresponding formulae for $\mu_j$ and $\nu_j$ follow from
(\ref{eq:28}) as real and imaginary parts. It is not
difficult to derive also the equations for the rest two
parameters $\xi_j$ and $\delta_j$ generalizing those
(\ref{eq:23}) and (\ref{eq:24}) for the two-soliton interaction.
Keeping in mind, however, our aim to formulate the equations for
the adiabatic interaction of the MNSE solitons in the form
tractable analytically, it is sufficient to represent the
equations for $\xi_j$ and $\delta_j$ in the following form:
\begin{eqnarray}
\frac{\mbox{d}\xi_j}{\mbox{d}t}&=&-\frac{\mu_j}{\alpha}+{\cal O}(\epsilon),
\label{eq:29}\\
\frac{\mbox{d}\delta_j}{\mbox{d}t}&=&\frac{\mu_j^2+\nu_j^2}{2\alpha^2}+
{\cal O}(\epsilon).\label{eq:30}
\end{eqnarray}
Indeed, let us impose the conditions on the scattering data of the
spectral problem (\ref{eq:2}) which correspond to the adiabatic
approximation. Just as for the NSE, we require that the eigenvalues of the
Lax operator are clustered around their mean value:
\[
|\lambda _j -\lambda _0|^2 \simeq {\cal  O}(\epsilon ), \qquad
\lambda _0 = {1  \over N } \sum_{j=1}^{N} \lambda _j.
\]
Thus we obtain the estimates in (\ref{eq:29}) and (\ref{eq:30}) which
mean that we can neglect the perturbation-induced evolution of the
parameters $\xi_j$ and $\delta_j$ as compared to their main
(unperturbed) evolution. At the same time $s_j $ and $w_j $
characterize the initial conditions and it is important to take
them into account in the right hand side of the equation (\ref{eq:28}).

\section{Derivation of the complex Toda chain model}\label{sec:4}

The next important step towards deriving a model of $N$ MNSE-soliton
interactions tractable analytically consists in a careful account for the
terms of the order of $\epsilon$. First note that because the right-hand
side of (\ref{eq:28}) is of the order of $\epsilon$, we may approximate
$k_j$ by $|k_0|e^{-is_j} $ where $k_0 $ is the mean value
\[
k_0=\frac{1}{N}\sum_{j=1}^Nk_j.
\]
Thereby we neglect terms like $|\nu_0-\nu_j|\epsilon$ and $|\mu_0-\mu_j|
\epsilon$ which due to (\ref{eq:15}) and (\ref{eq:18}) are of the higher
order in $\epsilon$. Hence, eq. (\ref{eq:28}) is written as follows:
\begin{equation}
\frac{\mbox{d}\lambda_j}{\mbox{d}t}=\frac{4\nu_0^3}{\alpha^2}
\left( e^{Q_{j+1}-Q_j}f_j - e^{Q_j-Q_{j-1}} g_j\right), \label{eq:31}
\end{equation}
where
\begin{eqnarray}\label{eq:1.8}
Q_{j+1} - Q_j &=& - {\nu _0 \over \alpha } (\xi_{j+1} -\xi_j )
\nonumber\\
&-& i \left[ \pi + {\mu _0 \over \alpha } (\xi_{j+1} -\xi_j ) + \delta
_{j+1} - \delta _j + 4s_{j+1} + 4s_j \right], \\
\label{eq:1.8a}
f_j &=& e^{i(s_{j+1}-s_j)}, \qquad g_j = e^{i(s_{j-1}-s_j)}.
\end{eqnarray}
The recurrent relation (\ref{eq:1.8}) can be solved for $Q_j $ with the
result
\begin{equation}
Q_j=-\frac{\nu _0}{\alpha}\xi_j  -i \left[ j\pi + \frac{\mu_0}{\alpha}
\xi_j +\delta_j +\delta_0 +\sum_{k=1}^{j-1} 8s_k + 4s_j \right], \qquad
\delta_0=\frac{1}{N}\sum_{j=1}^N\delta_j.  \label{eq:32}
\end{equation}

The next step is to derive the evolution equation for $Q_j$ (\ref{eq:32}).
It should be noted first of all that up to terms of the order of
$\epsilon$
\begin{eqnarray}
\frac{\mbox{d}\delta_0}{\mbox{d}t}&=&\frac{1}{2\alpha^2N}
\sum_{j=1}^N(\mu_j^2+ \nu_j^2)\nonumber\\
&=&\frac{1}{2\alpha^2N} \sum_{j=1}^N\left[ \mu_0^2+\nu_0^2+ 2\mu_0
(\mu_j-\mu_0) +2\nu_0 (\nu_j-\nu_0)+{\cal O}(\epsilon)\right] \nonumber\\
&=&\frac{1}{2\alpha^2}\left(\mu_0^2+\nu_0^2+{\cal O}(\epsilon)\right).
\nonumber
\end{eqnarray}
Then, in view of (\ref{eq:29}) and (\ref{eq:30}), we get
\begin{eqnarray}\label{eq:33}
\frac{{\rm d}Q_j}{{\rm d}t}&=&\frac{i}{\alpha^2}(\mu_0-i\nu_0)
\mu_j-\frac{i}{2\alpha^2}(\mu_j^2+\nu_j^2+\mu_0^2 +\nu_0^2) \\
&=&\frac{i}{2\alpha^2}\left[-(\mu_j-\mu_0)^2-(\nu_j-\nu_0)^2-
2i\nu_0(\mu_j-i\nu_j)\right] \nonumber\\
&=&\frac{\nu_0}{\alpha^2}\lambda_j +{\cal O}(\epsilon). \nonumber
\end{eqnarray}
Finally, keeping only the leading-order terms and replacing $f_j \simeq 1$
and $g_j \simeq 1$  we find from (\ref{eq:1.8}) and (\ref{eq:33})
\begin{equation}
\frac{{\rm d}^2Q_j}{{\rm  d}t^2}=4\left(\frac{\nu_0}{\alpha}
\right)^4\left(e^{Q_{j+1}-Q_j}- e^{Q_j-Q_{j-1}}\right), \label{eq:34}
\end{equation}
i. e., the CTC model. Hence we see that the CTC model arises naturally as
the integrable limit of the Karpman-Solov'ev-like equations describing the
adiabatic interaction of $N$ MNSE solitons within the train of solitons
with near velocities and widths.

As we will see in the next section the interactions of the MNSE
solitons are substantially different from the ones of the NSE. As
it can be seen from (\ref{eq:1.8}) the dependence of $Q_j $ on
the soliton parameters is different from that for the NSE case. An
important point here is that $\im Q_j $ depends explicitly also
on the amplitudes of the solitons through $s_j =\arctan \left(
\nu _j/(\alpha (1+\mu _j)\right) $.

\section{Dynamical regimes of the $N $-soliton trains}\label{sec:5}

It is well known that the CTC is a completely integrable
dynamical system. Most of the results concerning the CTC such as
the Lax representation, the integrals of motion, explicit
solutions etc. are direct consequences of the classical results
by Toda and Moser~\cite{Toda,Moser} on the real Toda chain (RTC).
However there is a qualitative difference between the RTC and the
CTC when one tries to analyze the dynamical regimes of the two
systems, see \cite{Gerd1,Gerd3,GEI}.

Indeed, the Lax representation of the CTC (\ref{eq:34}) is of the
form:
\begin{eqnarray}\label{eq:Lax}
{dL  \over  d\tau } &=& [L, M], \\
\label{eq:L} L(\tau)&=&\sum_{j=1}^{N} \left( b_j E_{j,j} +
a_j(E_{j,j+1}+ E_{j+1,j})
\right), \\
\label{eq:M} M(\tau) &=& \sum_{j=1}^N a_j(E_{j+1,j} - E_{j,j+1}),
\end{eqnarray}
where
\[
\tau = c_0 t, \qquad c_0=2\nu _0^2/\alpha ^2 ,\qquad a_j={1 \over
2}\exp{(Q_{j+1}-Q_j)/2} ,
\]
\[
b_j=-{1\over 2} { {\rm d}Q_j/{\rm d}\tau }=-\lambda_j/4\nu_0,
\qquad (E_{i,j})_{ln} =\delta_{il} \delta _{jn} ,
\]
see (\ref{eq:33}). In fact, without loss of generality we can
assume that $\tr L=0 $. This can be achieved by subtracting
$\zeta _0\openone  $ from $L $ where $\zeta _0
=\sum_{j=1}^{N}\zeta _j/N = \sum_{j=1}^{N}b_j/N  $. Note that
$\zeta _0 $ is obviously an integral of motion for the CTC, i.
e., ${\rm d}\zeta _0/{\rm d}\tau =0 $.

The explicit solution to the CTC is given by
\begin{equation}\label{eq:qaa}
q_k(\tau)=q_1(0) + \ln {A_{k}(\tau)\over A_{k-1}(\tau)},
\end{equation}
where $A_0\equiv 1 $,
\begin{equation}\label{eq:A1}
A_1(\tau)=\sum_{k=1}^{N}r_{k}^{2} e^{-2\zeta_k \tau},
\end{equation}
\begin{eqnarray}\label{eq:Ak}
 A_{k}(\tau)= \sum_{1\leq l_{1}< \ldots <l_{k}\leq N}
(r_{l_1} \ldots r_{l_k})^{2} W^{2}(l_k,l_{k-1},\ldots ,l_1)
e^{-2(\zeta _{l_1} + \dots + \zeta _{l_k})\tau }
\end{eqnarray}
and
\begin{equation}\label{eq:A_NConst}
A_N =  W^2(N,N-1,\dots,2,1) e^{-2(\zeta _{1} + \dots + \zeta
_{N})\tau } \prod_{k=1}^{N} r_k^2.
\end{equation}
Here $\zeta _j $ are the eigenvalues of the Lax matrix $L $,
$W(l_k,\dots, l_1) $ denotes the Vandermonde determinant:
\begin{equation}\label{eq:Wan}
W(l_k,\dots , l_1) = \prod_{\scriptsize \begin{array}{c} {s>p} \\
{s,p \in \{ l_1,\dots , l_k\}}\end{array}} (2\zeta _s - 2\zeta
_p),
\end{equation}
and $r_j $ are the first components $r_j=\vec{v}_{j,1} $ of the
eigenvectors:
\begin{equation}\label{eq:L-ev}
L \vec{v}_j = \zeta _j \vec{v}_j,
\end{equation}
normalized by
\begin{equation}\label{eq:v-norm}
(\vec{v}_j, \vec{v}_j) \equiv \sum_{k=1}^{N} \left( \vec{v}_{j,k}
\right)^2 =1.
\end{equation}
Due to the fact that $L $ is a symmetric matrix we find also
\begin{equation}\label{eq:r-norm}
\sum_{j=1}^{N} r_j^2 = 1.
\end{equation}

Using the explicit solution for $Q_j(t) $ we can estimate the asymptotic
behavior of $Q_j(\tau ) $ for $\tau \to \infty  $.

Such an analysis for the RTC, i. e., when $Q_j $, $a_j $, and $b_j
$ are real, shows that: i)~$r_j $ and $\zeta _j $ are real-valued,
ii)~$\zeta _j\neq \zeta _k $ for $j\neq k $. Therefore one finds
that for $\tau\to\infty  $ each  'particle' $Q_j $ moves
uniformly with a velocity $2\zeta _j $ \cite{Moser}. Since $
\zeta _j $ are pair-wise different we conclude that the only
possible dynamical regime is the asymptotically free (AFR) one.

The same considerations applied to the CTC lead however to a
qualitatively different results. Indeed, now $r_j $ and $\zeta_j =\kappa
_j+i\eta_j $ become complex-valued and there are no restrictions on the
eigenvalues $\zeta _j $. Then evaluating the limits of $Q_j(\tau ) $ for
$\tau \to\infty  $ we find that the asymptotic velocity of $Q_j $ is
determined by $2\kappa _j = 2\re \zeta _j$. As a result we have a much
wider spectra of dynamical regimes. The reason for that is also in the
fact that CTC can be viewed as a dynamical system of $N $ `complex`
particles which are characterized not only by their positions $\re Q_j $
and velocities $v_j=\re b_j $, but also by their phases and phase
velocities; the latter are related to $\im Q_j $ and $\im b_j $.
Physically speaking these `complex` particles have, just like the bright
NLS solitons, an internal degree of freedom. This makes the interaction
between the particles more complicated and as a result the number of the
possible dynamical regimes increases substantially.

The AFR  which takes place if $\kappa _j\neq \kappa _k $ for
$j\neq k $ is just one of the options. Another option is $\kappa
_1=\kappa _2 = \dots =\kappa _N =0$ which corresponds to a bound
state regime (BSR) of all $N $ `complex' particles (solitons) in
the train. There is also a large class of intermediate or mixed
regimes (MR) for which only several of the parameters $\kappa _j
$ are equal. For example, if $\kappa _1=\kappa _2=\kappa _3 >
\kappa _4 \dots >\kappa _N $ then the first three particles
(solitons) will form a bound state while the rest $N-3 $
particles will be asymptotically free.

Note that this variety of regimes exist in the generic case when
the eigenvalues $\zeta _j $ of $L $ are pair-wise different; so in
the previous case we assume that $\eta_1\neq \eta_2 \neq \eta_3
$. One may consider also degenerate regimes (when two or more of
the eigenvalues $\zeta _j $ become equal) and singular regimes
(when one or more of the functions $Q_j(\tau ) $ develop
singularities for finite $\tau  $).

There is also another important consequence from the
integrability of CTC. From the Lax representation one easily
finds that the eigenvalues $\zeta _j $ are the integrals of
motion for the CTC, i.e., $\zeta _j $ are time independent.
Therefore we can evaluate them, for example, at the initial moment
$t =0 $ using for this the initial values of the soliton
parameters. Then, knowing $\zeta _j $  and, more specifically,
$\kappa _j $ we can predict the asymptotic regime of the
corresponding $N $-soliton train.

We can also answer another question: describe the set of initial
soliton parameters for which the corresponding $N $-soliton train
will develop a specific dynamic regime. In other words, we can
describe the set of initial soliton parameters for which we will
have, say, an $N $-soliton bound state regime. To describe the
BSR all we need to do is to solve the corresponding
characteristic equation
\begin{equation}\label{eq:ch-eq}
\det (L-\zeta \openone ) =0,
\end{equation}
and impose the condition $\kappa _1=\kappa _1=\dots=\kappa _N=0 $.
Since the coefficients of (\ref{eq:ch-eq}) and consequently
$\kappa _j $ will be expressed in terms of the initial soliton
parameters we will have a set of nonlinear equations describing
the BSR. Analogously, if we need to describe the AFR we must solve
for $\kappa _j\neq \kappa _k $ for $k\neq j $.

We will show how this can be done analytically for the simplest
non-trivial cases with $N=2 $ and $N=3 $. For generic values of
$N $ this can always be done by numeric means; one needs only to
solve  algebraic equation (\ref{eq:ch-eq}) of order $N $.

Let us briefly describe the manifolds of soliton parameters
responsible for each  of the dynamical regimes for $N=2 $
and $N=3 $. As it is clear from the above considerations, we have
to solve the characteristic equation (\ref{eq:ch-eq}) and to
express the eigenvalues $\zeta _j $ of $L $ in terms of the
soliton parameters.

\subsection{$ N=2 $ case.}\label{ssec:5.1}

For simplicity from now on we shall consider trains with zero
initial velocities, $\mu{_j(0)}=0$, i. e., in the relevant moving
coordinate system. The matrix $L_0\equiv L(t=0)=\pmatrix{b &a\cr
a &-b}$ with $\tr L=0 $ is built from the initial soliton
parameters:
\[
a=-\frac{i}{2}\exp\left(-\frac{\nu_0}{2\alpha}r_0-\frac{i}{2}\Gamma\right),
\qquad b=\frac{i}{4}d,
\]
where
\begin{equation}\label{eq:5.5}
r_0=\xi_{2(0)}-\xi_{1(0)}, \qquad \Gamma=\delta_{2(0)}- \delta_{1(0)}+
4s_1+ 4s_2, \qquad d=(\nu_{1(0)}-\nu_0)/\nu_0.
\end{equation}
Then
\begin{equation}\label{eq:6.2}
\zeta _{1,2} = \pm \sqrt{b^2 + a^2} = \pm {i \Delta_{\rm cr,2}
\over 4} \sqrt{y_0^2 + e^{-i\Gamma}}.
\end{equation}
with
\begin{equation}\label{eq:5.5a}
\Delta _{\rm cr,2} = 2e^{-\nu _0r_0/(2\alpha) }, \qquad y_0 = {d
\over \Delta _{\rm cr,2} }.
\end{equation}
Obviously if $\Gamma  \neq 0, \pi $ then $\re \zeta _{1,2} \neq 0
$ and we will have an asymptotically free regime (AFR).

If $\Gamma =0 $, then $\re \zeta _{1,2}=0 $ and we have a bound
state regime (BSR).

If $\Gamma =\pi $, then $\re \zeta _{1,2}=0 $, i. e., we will have
a BSR only provided
\begin{equation}\label{eq:6.3}
|d| > \Delta _{\rm cr,2}.
\end{equation}
If $|d| < \Delta _{\rm cr,2} $, both roots $\zeta _{1,2} $ become
real and we go into the AFR.

It was already noted that the conditions $\Gamma  =0, \pi $
involve, besides the phases $\delta _j $, also the amplitudes of
the solitons through $s_j $. In particular for $\nu _0=\alpha =1
$ and $\mu _j=0 $ we have $s_1 + s_2 = \pi/4 $. Therefore two
such MNSE solitons {\em attract} each other and form a bound state
provided $\delta _2-\delta _1=\pi $ and {\em repulse} each other
(which leads to  AFR) for $\delta _2-\delta _1=0 $. Such behavior
of the two-soliton interaction is quite to the contrary to that
known for the NSE two-soliton interaction.

The explicit solution to the CTC with $N=2 $ is of the form
\begin{eqnarray}\label{eq:6.4}
Q_1(t) &=& -Q_2(t) = \ln {\cosh(2\zeta _1 c_0 t -\gamma _1)  \over 2\zeta
_1 }, \qquad \gamma _1 = \ln {r_1  \over r_2 },
\end{eqnarray}
where
$\zeta _1 $ is expressed in terms of the soliton parameters
(\ref{eq:6.2}) and
\begin{eqnarray}\label{eq:9.1}
\gamma_1 = {1 \over 2 } \ln {\sqrt{y_0^2 +
e^{-i\Gamma }} + y_0 \over \sqrt{y_0^2 + e^{-i\Gamma }} - y_0 }.
\end{eqnarray}
Obviously for $\Gamma =0 $ the solution $Q_1(t) $
\begin{eqnarray}\label{eq:55'}
Q_1(t) &=& \ln {2\cos (Y_0 c_0 t/2 + i\gamma _{10}  \over iY_0 }, \\
Y_0 &=& \Delta _{\rm cr,2} \sqrt{y_0^2+1}, \qquad \gamma _{10} =
{1  \over 2 } \ln {\sqrt{y_0^2 +1} + y_0  \over \sqrt{y_0^2 +1} - y_0 }.
\nonumber
\end{eqnarray}
becomes a periodic function of $t=\tau/c_0 $ with period depending on
$y_0 $:
\begin{equation}\label{eq:9.5}
T_{\rm 2s;1} = {4\pi \over c_0\Delta _{\rm cr,2} \sqrt{y_0^2 +1} }.
\end{equation}

Analogously for $\Gamma =\pi $ from (\ref{eq:9.1}) we have
\begin{eqnarray}\label{eq:9.6}
Q_1(t) &=& -Q_2(t) = \ln {2 \cosh(i\Delta _{\rm cr,2}\sqrt{y_0^2 -1} c_0
t/2 -\gamma _{11}) \over i\Delta _{\rm cr,2} \sqrt{y_0^2-1} },
\nonumber\\
\gamma _{11} &=& {1\over 2} \ln {\sqrt{y_0^2-1} +y_0 \over \sqrt{y_0^2-1}
-y_0 }.
\end{eqnarray}
The solution is periodic only if $y_0>1 $ and the period is
\begin{equation}\label{eq:9.7}
T_{\rm 2s;2} = {4\pi  \over c_0\Delta _{\rm cr,2}\sqrt{y_0^2-1} }.
\end{equation}
As a conclusion, the BSR for $N=2 $  provides periodic
solutions. For $\Gamma =\pi $, $y_0<1 $ we have AFR and the
solution is not periodic.

The final remark in this subsection is that for $y_0\to 0 $ the
solution (\ref{eq:6.4}) becomes singular and blows up
periodically with period $4\pi/(c_0\Delta _{\rm cr,2}) $. In this
limit we have two `equal' solitons with amplitudes $\nu _j=\nu
_0=1 $ with phase difference $\pi $.

\subsection{$ N=3 $ case.}\label{ssec:5.2}

For the case of the three-soliton train with zero initial
velocities the matrix $L_0$ has the form
\[
L_0=\pmatrix{b_1 &a_1& 0\cr a_1 &b_2 &a_2\cr 0& a_2& b_3}, \qquad
\tr L_0=0
\] with
\[
a_j=-{i\over
2}\exp(-\frac{\nu_0}{2\alpha}r_0-\frac{i}{2}\Gamma_j), \qquad
b_j={i\over4}d_j,
\]
where
\begin{eqnarray}\label{eq:N_3}
d_j &=& \frac{\nu_{j(0)}-\nu_0}{\nu_0}, \qquad
r_0=\xi_{2(0)}-\xi_{1(0)}=\xi_{3(0)}-\xi_{2(0)}, \nonumber\\
\Gamma_j &=& \delta_{j+1(0)}-\delta_{j(0)}+4s_{j+1}+4s_j.
\end{eqnarray}
Then the characteristic equation takes the form:
\begin{equation}\label{eq:16.1}
\zeta ^3 + p\zeta  + q=0,
\end{equation}
where
\begin{eqnarray}\label{eq:16.2}
p&=& - {1  \over 16 } (d_1 d_2 + d_2 d_3 + d_1 d_3 ) + {1  \over
4 } e^{-r_0\nu _0/\alpha } \left( e^{-i\Gamma _1} +
 e^{-i\Gamma _2} \right), \nonumber\\
q&=& {i  \over 64 } d_1d_2d_3 - {i  \over 16 } e^{-r_0\nu
_0/\alpha } \left(  d_1 e^{-i\Gamma _2} +  d_3 e^{-i\Gamma _1}
\right).
\end{eqnarray}

It is  natural to make use of the well known Cardano formulae for
solving cubic equations. We first consider the cases when $p $
and $q $ are real. The roots of (\ref{eq:16.1}) are given by
\begin{equation}\label{eq:13.2}
\zeta _1 = A+B, \qquad \zeta _2 = \omega A + \omega ^2 B, \qquad
\zeta _3 = \omega^2 A + \omega B,
\end{equation}
where
\begin{eqnarray}\label{eq:13.3}
A &=& \sqrt[3]{-{ q \over 2 } + \sqrt{Q}}, \qquad
B = \sqrt[3]{-{ q \over 2 } - \sqrt{Q}}, \\
Q&=& {q^2  \over 4 } + {p^3  \over 27}, \qquad \omega =
\exp\left(\frac{2\pi i}{3}\right). \nonumber
\end{eqnarray}

If both $p $ and $q $ are real, then so is $Q $. Here we have four
subcases corresponding to qualitatively different sets of roots
for real $p$ and $q$.

{\bf i)} ${\bf Q < 0 }$. This is possible only if $p< p_{\rm cr}
$, $p_{\rm cr}=-3(q^2/4)^{1/3} $. Then $A=B^* $ and all three
roots $\zeta _j $ become real $\zeta _j=\kappa _j $ and pair-wise
different:
\begin{equation}\label{eq:AFR}
\kappa _1 = 2 |A|\cos \Omega _0, \qquad \kappa _{2,3} = 2 |A|\cos
\left(\Omega _0 \pm {2\pi \over 3 }\right),
\end{equation}
with $\Omega _0\neq 0, \pi $. Obviously this leads to  AFR. If
$\Omega _0=0 $ or $\pi $ then $\kappa _2=\kappa _3 $ and a MR
follows.

{\bf ii) ${\bf Q>0} $ and ${\bf q\neq 0} $.} Here both $A $ and
$B $ are real and formula (\ref{eq:13.2}) shows that one root
$\zeta_1$ is real, while the other two are complex conjugate:
\begin{equation}\label{eq:MR}
\re \zeta _1 = - 2\re \zeta _2 =-2\re \zeta _3, \qquad \mbox{or}
\qquad \kappa _1 = - 2\kappa _2 =-2\kappa _3,
\end{equation}
which corresponds to a MR.

{\bf iii) ${\bf Q>0} $ and ${\bf q= 0} $.} Now $p>0$, the cubic
equation (\ref{eq:16.1}) simplifies and is trivially solved by
\begin{equation}\label{eq:BSR-r}
\zeta _1=0, \qquad \zeta _{2,3} =\pm \sqrt{-p}.
\end{equation}
All the roots have zero real parts which obviously corresponds to
BSR.

{\bf iv) ${\bf Q=0} $}. If $p$ and $q$ are nonzero, all the rots
are real and pair-wise different:
\[
\zeta_1=3q/p, \qquad \zeta_2=\zeta_3=-3q/2p,
\]
we have AFR. If $p$ and $q$ are zero, we get a degenerate case
with all three zero roots.

The symmetry in the eigenvalues leads also to a symmetry in the
solutions of the CTC. Therefore the configuration
(\ref{eq:BSR-r}) corresponds to a particular type of BSR's. This
is due to the fact that we restricted so far both $q $ and $p $ to
be real. Of course this is not necessary; moreover, from
(\ref{eq:16.2}) we see that generically both $q $ and $p $ are
complex. If we want to specify the soliton parameters that are
responsible for the BSR we may also use Viette formulae which show
that the characteristic equation (\ref{eq:16.1}) will have purely
imaginary roots if $p $ is real and negative and $q $ is purely
imaginary. That is why we will consider also the configuration
v) below.

{\bf v)~${\bf p={\bar p}} $, ${\bf q=-{\bar q}} $.} In this case
we have two qualitatively different possibilities depending on
whether $Q $ is positive or negative.

Note that since $q=-{\bar q} $ we should modify our reasoning as
compared to the above analysis. Indeed, with $q=iq' $, $q' $ real
and $Q\geq 0 $ we find that $A=-\bar B $. Therefore from
(\ref{eq:13.2}) and (\ref{eq:13.3}) we have that all the roots
$\zeta _k $ satisfy $\zeta _k=-\bar\zeta _k $, i. e., are purely
imaginary and BSR takes place.

Analogously, if $Q<0 $ then the roots $\zeta _k $ satisfy $\zeta
_1=-\bar\zeta _1 $ and $\zeta _3=-\bar\zeta _2 $ which leads to
AFR.

Hence, we revealed two possibilities to realize bound state
regime: subcase iii) and subcase v) with $Q>0$.

Let us now briefly describe the sets of soliton parameters relevant to
each of the regimes mentioned above. For definiteness we will use two
configurations of soliton widths:
\begin{eqnarray}\label{eq:23.2a}
\mbox{W1:} \qquad d_1=-d_3, \qquad d_2=0, \\
\label{eq:23.2b} \mbox{W2:} \qquad d_1=d_3, \qquad d_2=-2d_1.
\end{eqnarray}

The condition that $p $ is real immediately means that
\begin{equation}\label{eq:23.3}
\Gamma _1=-\Gamma _2\equiv\Phi .
\end{equation}
Then
\begin{eqnarray}\label{eq:24.1}
p&=& -{1  \over 16 } (d_1 d_2 + d_1d_3 + d_2 d_3) + {1  \over 2 } e^{-\nu
_0r_0/\alpha } \cos \Phi , \\
\label{eq:24.2}
q&=& {i  \over 64 } d_1 d_2 d_3 - {i \over 16 } e^{-\nu _0r_0/\alpha }
\left( d_1 e^{i\Phi } + d_3 e^{-i\Phi } \right), \\
\label{eq:24.3} \Phi  &=& \delta _2 - \delta _1 + 4s_1 + 4s_2.
\end{eqnarray}
Choosing the sets of widths to be W1 and W2 we get respectively:
\begin{eqnarray}\label{eq:24.4a}
p^{(1)} &=& {d_1^2  \over 16 } + {\epsilon _0^2  \over 2 } \cos
\Phi ,
\nonumber\\
q^{(1)} &=&  {d_1\epsilon _0^2  \over 8 } \sin \Phi ,
\end{eqnarray}
where $\epsilon _0 = \exp(-\nu _0r_0/(2\alpha )) $, and
\begin{eqnarray}\label{eq:24.4b}
p^{(2)} &=& {3d_1^2  \over 16 } + {\epsilon _0^2  \over 2 } \cos
\Phi ,
\nonumber\\
q^{(2)} &=& -{id_1^3 \over 32 } -{id_1\epsilon _0^2  \over 8 } \cos
\Phi.
\end{eqnarray}

{\bf Case I.} ${\bf q=0 }$.  The characteristic equation
(\ref{eq:16.1}) has the roots
\begin{equation}\label{eq:24.5}
\zeta _1=0, \qquad \zeta _{2,3} =\pm \sqrt{-p}.
\end{equation}
From (\ref{eq:24.4a}) we get that for the W1 configuration the
condition $q^{(1)}=0 $ holds provided
\begin{equation}\label{eq:24.6}
\Phi  =k\pi, \qquad k=0,1,
\end{equation}
which means that
\begin{equation}\label{eq:24.7}
p^{(1)} = {d_1^2  \over  16} + (-1)^k {\epsilon _0^2  \over 2 }.
\end{equation}
As a consequence we find that $p^{(1)}>0 $ for $k=0 $; for $k=1 $ we get
that $p^{(1)}>0 $ only provided $|d_1| $ is greater than the critical
value:
\begin{equation}\label{eq:24.8}
|d_1| > \Delta _{\rm cr,3} , \qquad \Delta _{\rm cr,3} = 2 \sqrt{2}
\epsilon _0.
\end{equation}
In all these cases $\zeta _{2,3} $ are purely imaginary, i.e.
these sets of parameters lead to  BSR.

Note that (\ref{eq:24.6}) means
\begin{equation}\label{eq:24.9}
\delta _2 = \delta _1 + k\pi - 4s_1 - 4 s_2, \qquad k=0,1.
\end{equation}

If instead of (\ref{eq:24.8}) we have $|d_1|<\Delta _{\rm cr,3} $
then $p^{(1)}<0 $ and the roots $\zeta _{2,3} $ become real. That
means that taking $d_1 $ below the critical value we will see a
transition from  BSR to  AFR.

The same considerations applied to the W2 configuration lead to
different results. From (\ref{eq:24.4b}) we see that $q^{(2)}=0 $
holds if
\begin{equation}\label{eq:25.1}
\cos \Phi  = - {d_1^2  \over 4\epsilon _0^2 },
\end{equation}
which implies that
\begin{equation}\label{eq:25.2}
|d_1| \leq 2\epsilon _0 = {\Delta _{\rm cr,3}  \over \sqrt{2} }
\end{equation}
and
\begin{equation}\label{eq:25.3}
p^{(2)} = {d_1^2  \over 16 } \geq 0.
\end{equation}
Such configurations obviously lead to BSR. If $|d_1| $ is chosen
to be greater than the critical value in the right hand side of
(\ref{eq:25.2}) we find that then $q^{(2)} $ becomes purely
imaginary; such situation is considered below.

Let us briefly treat also the case of `equal' solitons, i.e.,
$d_j=0 $. Then obviously $q=0 $, $s_1=s_2=s_3=\pi/8 $ and $
p=(\epsilon _0^2/2) \cos \Phi$. As a result we find that if
\begin{equation}\label{eq:25.7}
- {\pi  \over 2 } < \Phi  < {\pi  \over 2 }, \qquad \mbox{i.e.,}
\qquad {\pi  \over 2 } < \delta _2 - \delta _1 < {3\pi  \over 2 },
\end{equation}
then $p>0 $ and we have  BSR; if
\begin{equation}\label{eq:25.8}
{\pi  \over 2 } < \Phi  < {3\pi  \over 2 }, \qquad \mbox{i.e.,}
\qquad -{\pi  \over 2 } < \delta _2 - \delta _1 < {\pi  \over 2 },
\end{equation}
then $p<0 $ and  AFR follows.

{\bf Case II. ${\bf p=0} $.} In this case the characteristic
equation (\ref{eq:16.1}) has as roots
\begin{equation}\label{eq:25.4}
\zeta _k = \sqrt[3]{-q} \omega ^k, \qquad \omega =e^{2\pi i/3},
\qquad  k=0,1,2.
\end{equation}
If in addition $q $ is real then (\ref{eq:25.4}) leads to a MR;
otherwise we get  AFR.

For the W1 configuration $p^{(1)}=0 $ means
\begin{equation}\label{eq:25.5}
\cos \Phi  = - {d_1^2  \over (\Delta _{\rm cr,3})^2 };
\end{equation}
this is possible only if $|d_1| \leq \Delta _{\rm cr,3} $. From
(\ref{eq:24.4a}) we get that $q^{(1)} $ is real and such
configuration leads to MR.

For the W2 configuration $p^{(2)}=0 $ holds if
\begin{equation}\label{eq:25.6}
\cos \Phi = - {3d_1^2  \over 8\epsilon _0^2 } = - {3d_1^2  \over
(\Delta _{\rm cr,3})^2 },
\end{equation}
which is possible only if $|d_1| \leq \Delta _{\rm cr,3}/\sqrt{3}
$. From (\ref{eq:24.4b}) we find that $q^{(2)} $ is purely
imaginary, i.e.,  AFR follows.

{\bf Case III: ${\bf p={\bar p}}$ and ${\bf q=-{\bar q}\neq 0}$.}
This is possible only for the W2 configuration, so $p $ and $q $
are given by (\ref{eq:24.4b}). The resolvent of the cubic
equation (\ref{eq:16.1}) in this case takes the form:
\begin{eqnarray}\label{eq:26.1}
Q&=& {(p^{(2)})^3  \over 27 } + {(q^{(2)})^2  \over 4 }  \nonumber\\
&=& {\epsilon _0^6  \over 8 } \left[ \left(y^2 + {c  \over 3 }\right)^3 -
y^2 \left( y^2 + {c  \over 2 }\right)^2 \right] \nonumber\\
&=& {\epsilon _0^6  \over 8 } {c^2  \over 12 } \left(y^2 +
{4c\over 9}\right) ,
\end{eqnarray}
where $y= d_1/\Delta _{\rm cr,3} $ and $c=\cos \Phi $.

It is easy to check that $Q(y,c) $ is nonnegative for all $c>
-9y^2/4 $ and vanishes for $c=0 $ and $c=-9y^2/4 $. We have to
keep in mind also that $|c|\leq 1 $. Therefore if $9y^2/4 >1 $
then $Q\geq 0 $ in the whole interval $-1 \leq c \leq 1 $.
Following the arguments in v) above we conclude that this
configurations leads to BSR.

If we choose
\begin{equation}\label{eq:26.4}
|d_1| < {2  \over 3 }\Delta _{\rm cr,3}
\end{equation}
then there will be an interval for $\Phi $ (\ref{eq:24.3}),
\begin{equation}\label{eq:26.5}
\varphi _{\rm cr} \leq \Phi \leq 2\pi - \varphi _{\rm cr} , \qquad
\varphi _{\rm cr} = \arccos \left(-{9d_1^2 \over 4(\Delta _{\rm cr,3})^2
} \right) ,
\end{equation}
for which  $Q<0 $; i.e., if (\ref{eq:26.5}) holds we have AFR.

If $\Phi $ belongs to the complementary interval:
\begin{equation}\label{eq:26.6}
-\varphi _{\rm cr} \leq \Phi \leq \varphi _{\rm cr} ,
\end{equation}
then $Q(y,c)\geq 0 $ and we have BSR.

The interested reader can easily extend this studies to other relevant
configurations of soliton parameters.

\section{The CTC versus the numerical solutions of the MNLS}\label{sec:6}

It is our aim here to compare the predictions of the CTC model
with the numerical solutions of the MNSE.  Since  the full
numerical investigation of the problem is a voluminous and
ambitious task we limit ourselves with $N=2 $ and $N=3 $ soliton
trains and fix up $\alpha =1 $ and the average width $\nu _0=1 $.

With this choice of $\alpha =1 $ the derivative term in the MNSE
can not be treated as a perturbation to the NSE. With this choice
we are able to exhibit the differences between the MNSE and NSE
$N $-soliton train interactions. As we mentioned above the
dependence of the soliton interaction of the MNSE solitons on the
soliton phase difference is qualitatively different from the one
of the NSE solitons.

\begin{figure}{}
\epsfxsize=6.0cm
\epsfbox{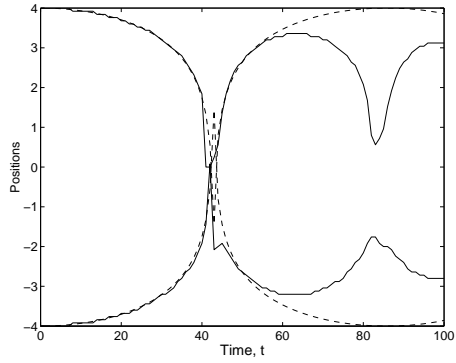}

\caption{Two-soliton interactions and their comparison with the CTC model.
Solid curve: numerical results; dashed curves: predictions from the
Toda chain model. $\nu _1=\nu _2=1.0 $, $\delta _1=0 $, $\delta _2=\pi $.
\label{fig:2s(f)}}
\end{figure}

Indeed, let us start with $N=2 $ soliton trains. The formulas from
Section~5.1  with $\alpha =1 $ and $\nu _0=1 $ show that `equal'
solitons (i.e., solitons with equal widths) with phase difference
$\delta _2-\delta _1=\pi $ (or $\Gamma =0 $) attract each other.  In fact
this choice of the soliton parameters corresponds to $y_0=0 $ and
according to (\ref{eq:6.4}), (\ref{eq:9.1}) the solution to the CTC becomes
singular. From Fig.~\ref{fig:2s(f)} we see that apart from a small
neighborhood around the singular points the CTC gives a good
description of the $2 $-soliton train of the MNSE; the singular
points match rather well with the points at which the two
solitons are closest to each other. The distance to the first
singular points matches $T_{\rm 2s,1}/4 $ with $T_{\rm 2s,1} $ given by
formula (\ref{eq:9.5}) with $y_0=0 $.

\begin{figure}{}
\epsfxsize=6.0cm
\epsfbox{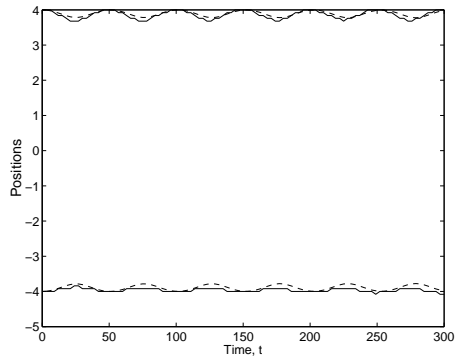}

\caption{Two-soliton interactions and their comparison with the CTC model.
Solid curve: numerical results; dashed curves: predictions from the Toda
chain model. $\nu _1= 0.95$, $\nu _2=1.05 $, $\delta _1=0 $, $\delta
_2=\pi $.\label{fig:2s(a)}}

\end{figure}

Choosing the solitons to have different widths leads to $\gamma
_{10}\neq 0 $ in  Eq.~(\ref{eq:55'}) and removes the singularity
of the corresponding solution of the CTC system even if $\Gamma
=0 $. This can be seen from Fig.~\ref{fig:2s(a)} which corresponds to a
BSR. Of course now the match between the MNSE simulation and the
CTC solution is better than in the previous case.

\begin{figure}{}
\epsfxsize=6.0cm
\epsfbox{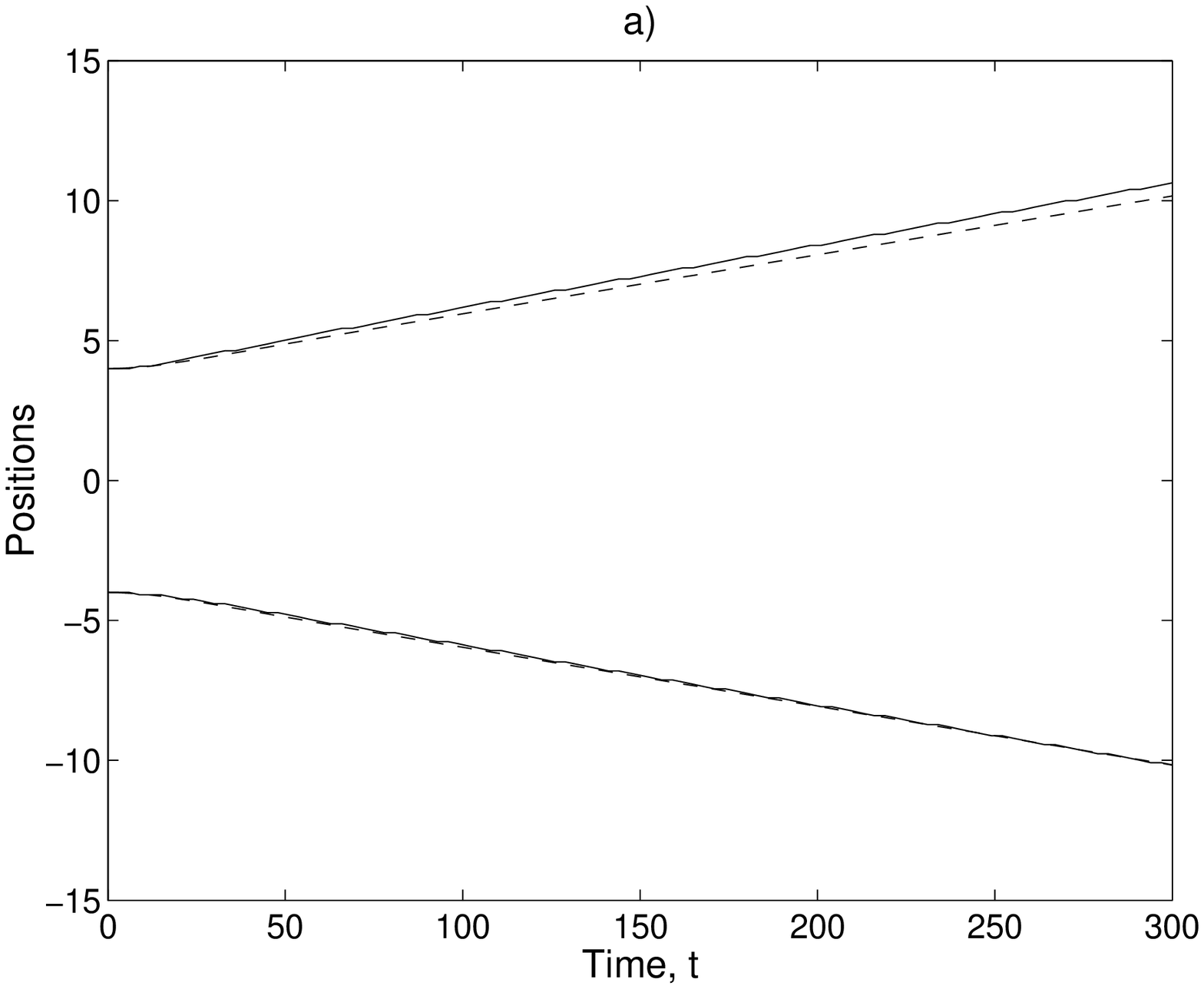}
\epsfxsize=6.0cm
\epsfbox{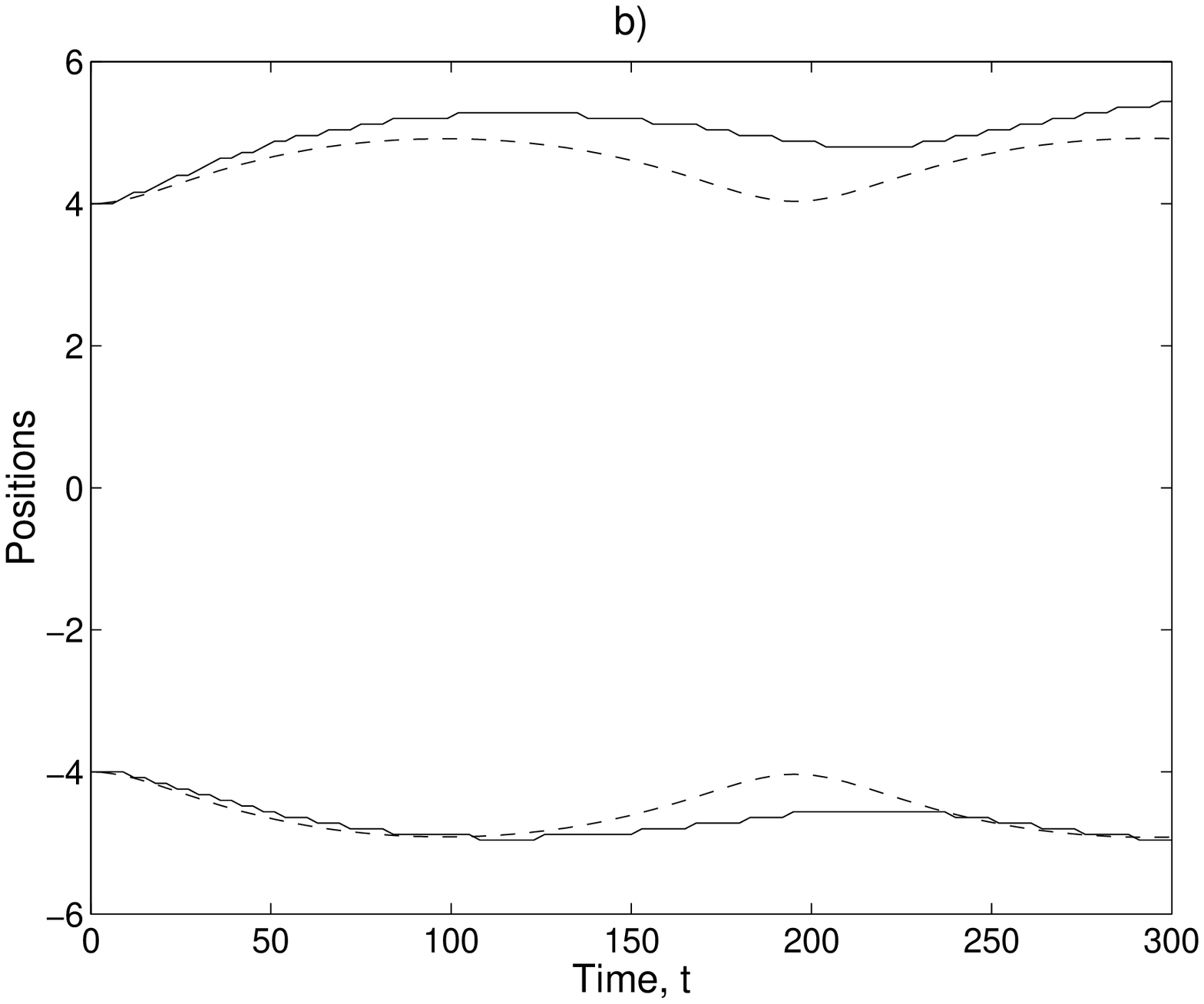}

\caption{Two-soliton interactions and their comparison with the CTC model.
Solid curve: numerical results; dashed curves: predictions from the Toda
chain model. a) $\nu _1= 0.97$, $\nu _2=1.03 $, $\delta _1=0 $, $\delta
_2=0 $; b) $\nu _1= 0.96$, $\nu _2=1.04 $, $\delta _1=0 $, $\delta
_2=0 $.\label{fig:2s(c-b)} }

\end{figure}

The situation changes if we consider solitons with phase
differences such that $\Gamma =\pi $. There we find a threshold
value for $d_1=-d_2=(\nu_1-\nu_0)/\nu _0 $, see
Eq.~(\ref{eq:6.3}). Whenever $d_1<\Delta _{\rm cr,2} $ we get an
AFR (see Fig.~\ref{fig:2s(c-b)}a while for $d_1>\Delta _{\rm cr,2} $ we get
an BSR (see Fig.~\ref{fig:2s(c-b)}b.)

Let us now consider the 3-soliton interactions. The choices of
the soliton parameters illustrates each of the three main
configurations outlined in Section~5.2 above.

\begin{figure}{}
\epsfxsize=6.0cm
\epsfbox{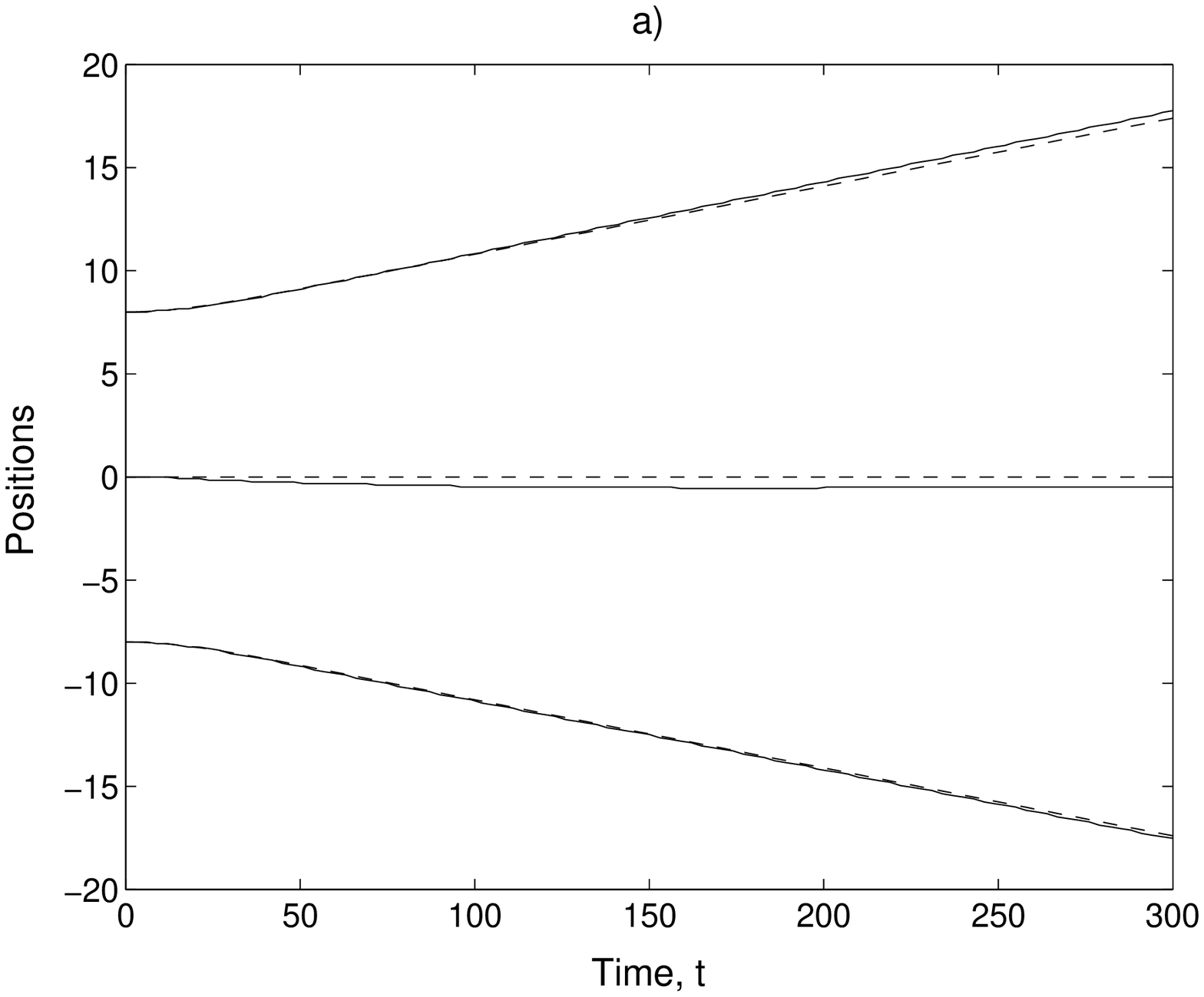}
\epsfxsize=6.0cm
\epsfbox{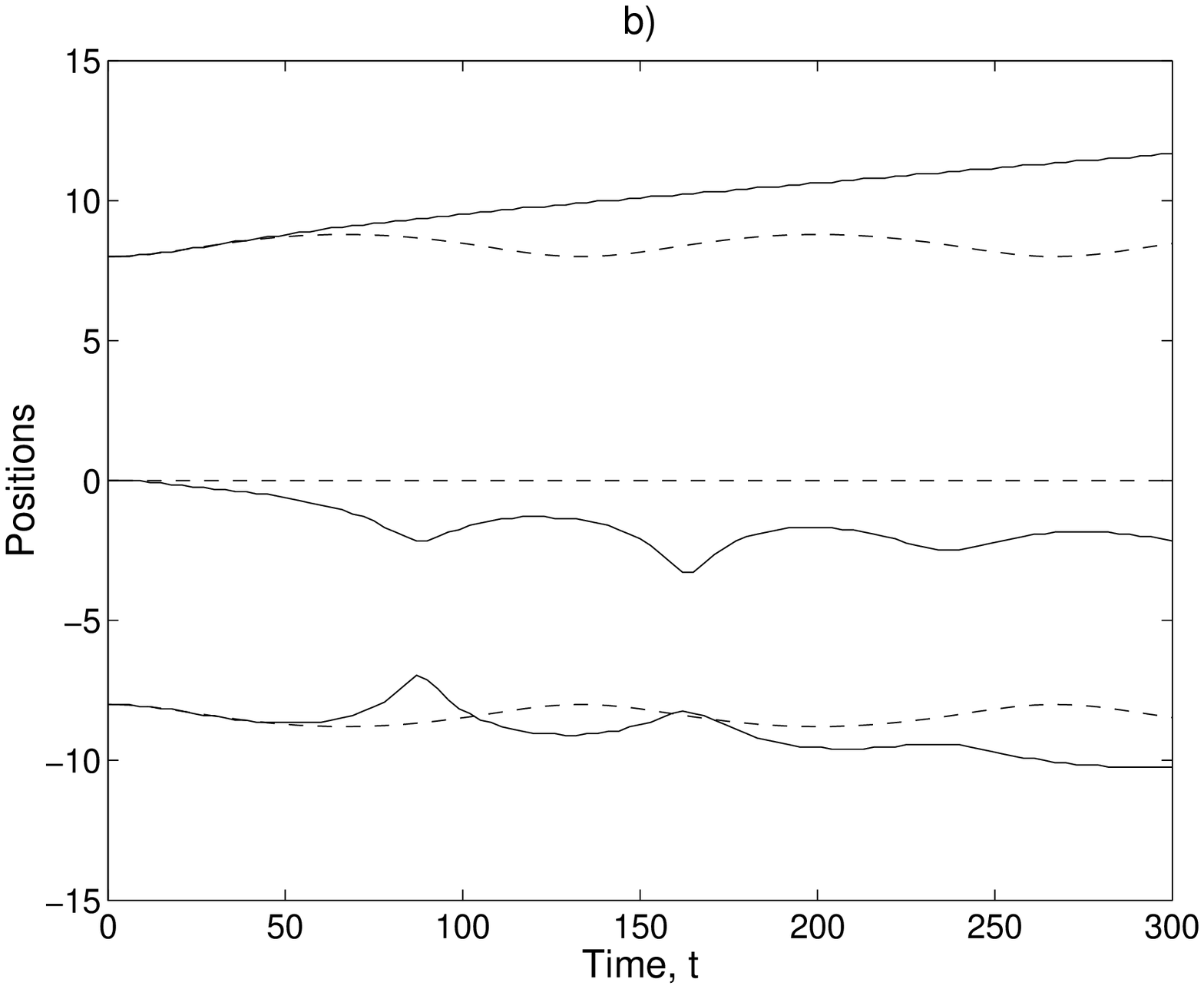}

\caption{Three-soliton interactions and their comparison with the CTC
model.  Solid curve: numerical results; dashed curves: predictions from
the Toda chain model. a) $\nu _1= 1.04$, $\nu _2=1.0 $, $\nu _3=0.96 $,
$\delta _1=0 $, $\delta _2=-0.0392 $, $\delta _3=0.0016 $; b) $\nu _1=
1.07$, $\nu _2=1.0 $, $\nu _3=0.93 $, $\delta _1=0 $, $\delta _2=-0.0676
$, $\delta _3=0.0049 $.\label{fig:Ib(i-ii)}}

\end{figure}

The figures \ref{fig:Ib(i-ii)} provide examples of 3-soliton
configurations with $q=0 $ characteristic for case I. Both sets
of parameters are such that $\Phi =\pi $. Besides on
Fig.~\ref{fig:Ib(i-ii)}a  we have $d_1<\Delta _{\rm cr,3} $ and as a
consequence an AFR must follow. In the next Fig.~\ref{fig:Ib(i-ii)}b we
have $d_1>\Delta _{\rm cr,3} $ for which the CTC model predicts a BSR; the
match with the simulation here is not that good.

\begin{figure}{}
\epsfxsize=6.0cm
\epsfbox{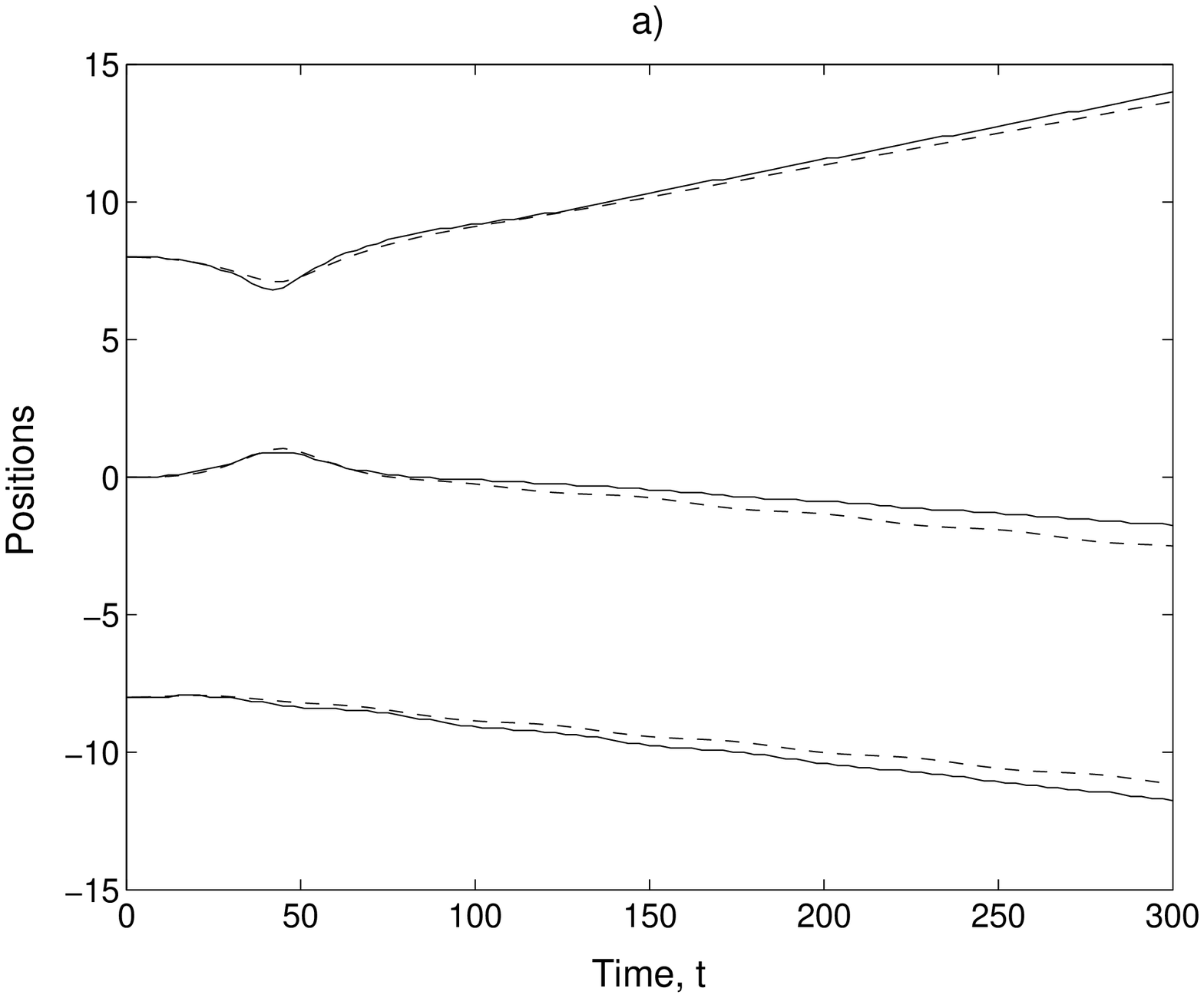}
\epsfxsize=6.0cm
\epsfbox{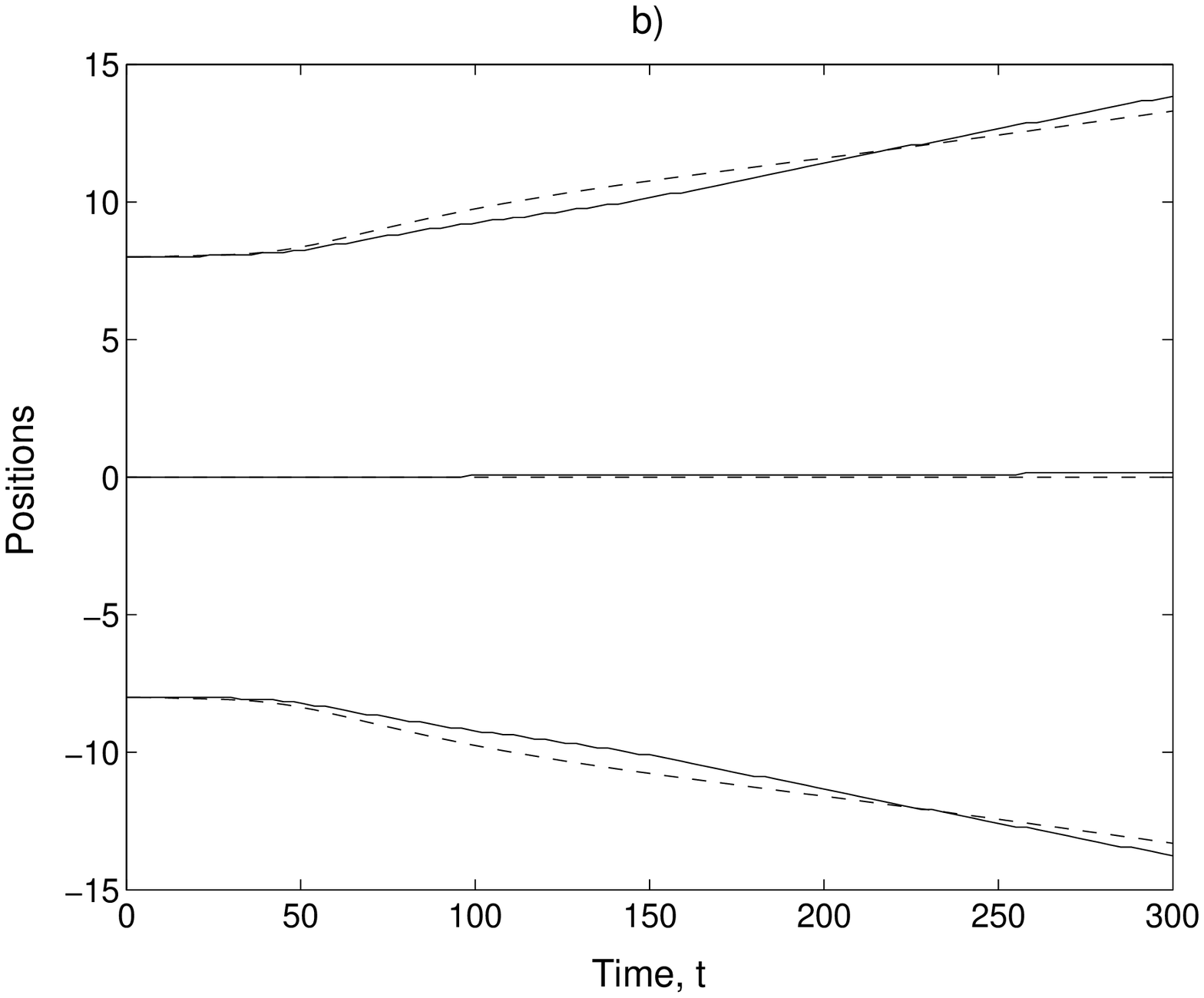}

\caption{Three-soliton interactions and their comparison with the CTC
model.  Solid curve: numerical results; dashed curves: predictions from
the Toda chain model. a) $\nu _1= 1.04$, $\nu _2=1.0 $, $\nu _3=0.96 $,
$\delta _1=0 $, $\delta _2=2.1703 $, $\delta _3=0.0016 $; b) $\nu _1=
1.02$, $\nu _2=0.96 $, $\nu _3=1.02 $, $\delta _1=0 $, $\delta _2=-1.0862
$, $\delta _3=0.0420 $.\label{fig:IIa(i)-b(ii)}}

\end{figure}

The figures \ref{fig:IIa(i)-b(ii)}  show a 3-soliton
configurations with $p=0 $ characteristic for case II. In
fig~\ref{fig:IIa(i)-b(ii)}b the set of widths is W1 and $d_1<\Delta _{\rm
cr,3} $ and therefore a MR follows

\begin{figure}{}
\epsfxsize=6.0cm
\epsfbox{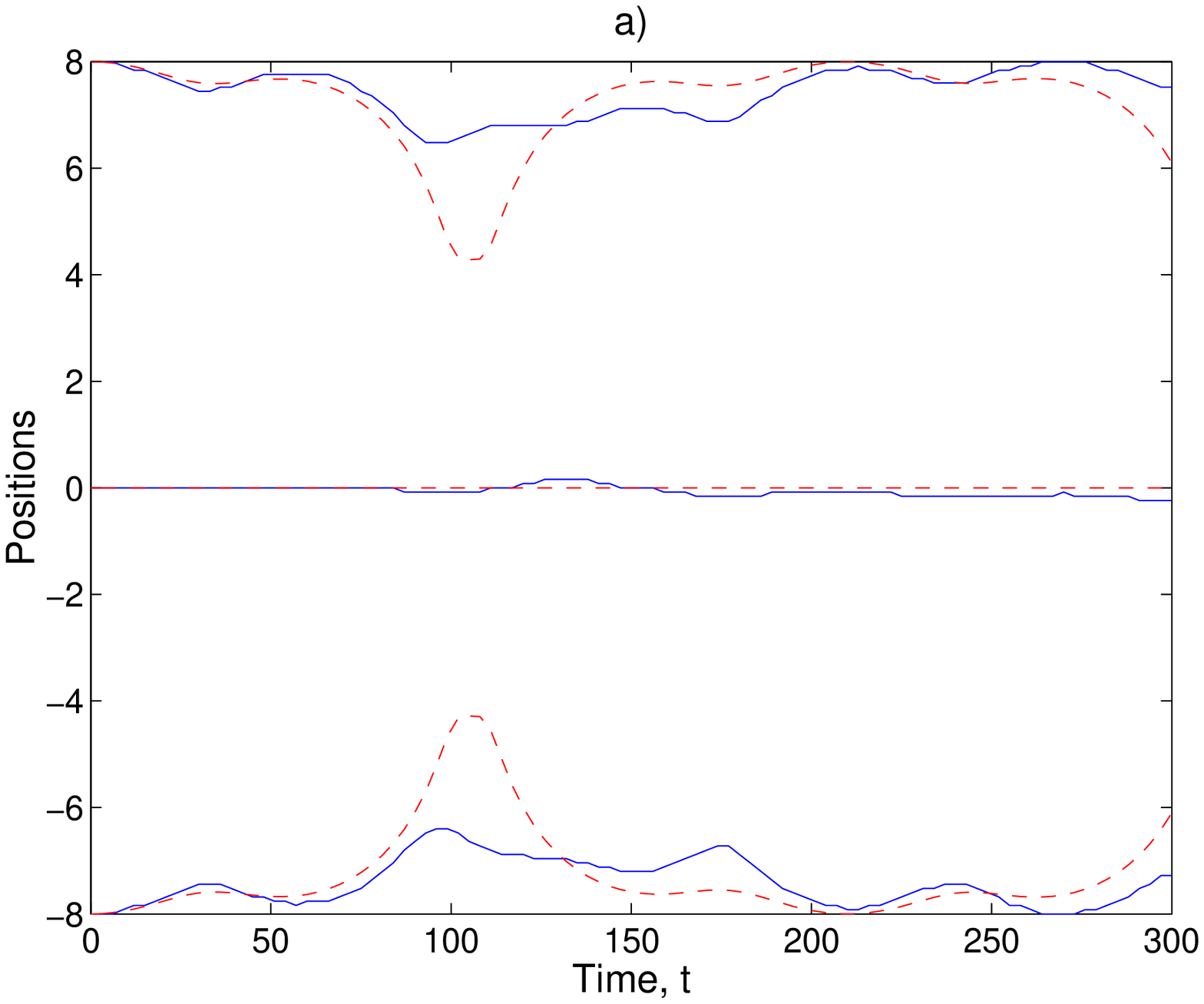}
\epsfxsize=6.0cm
\epsfbox{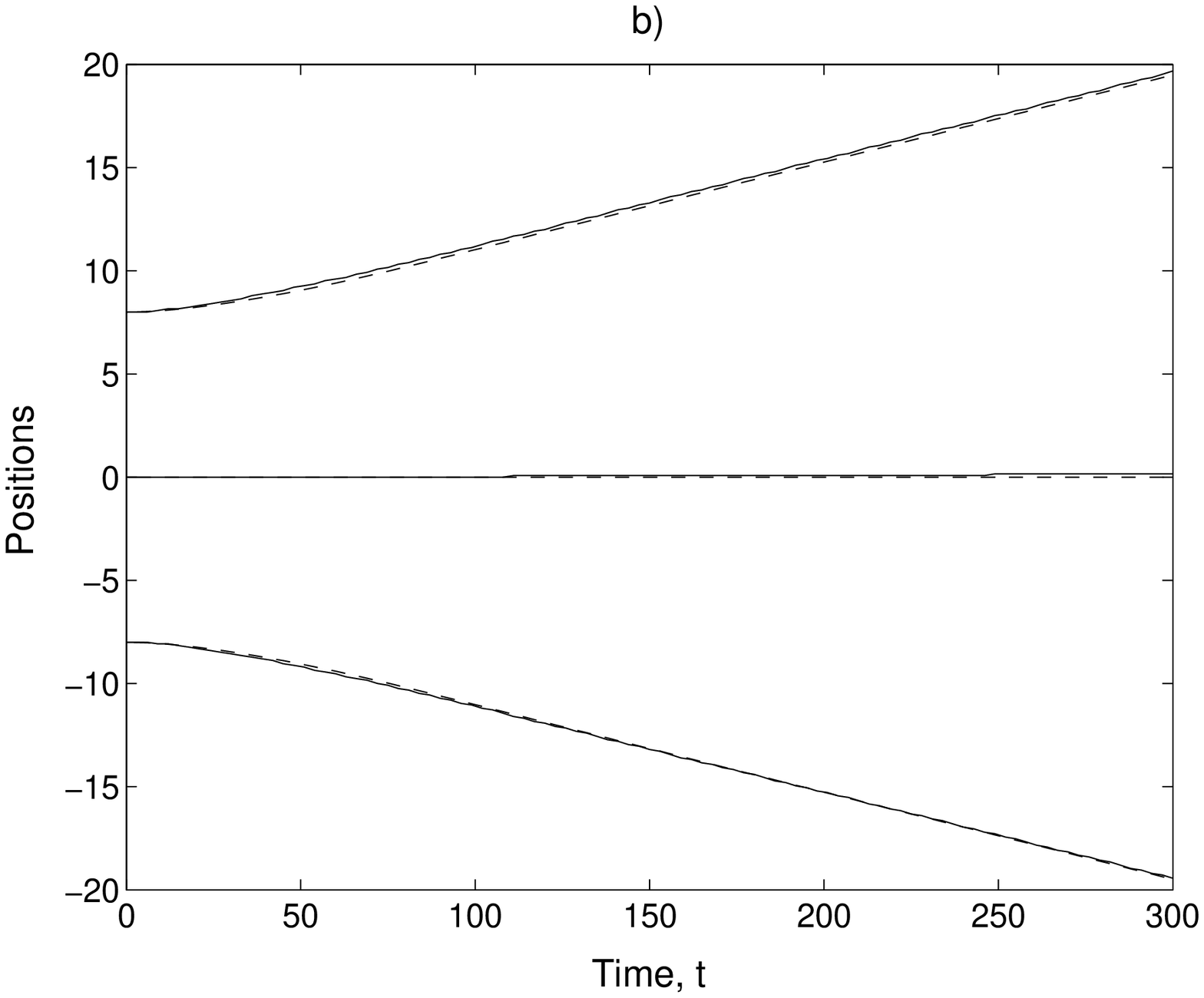}

\caption{Three-soliton interactions and their comparison with the CTC
model.  Solid curve: numerical results; dashed curves: predictions from
the Toda chain model. a) $\nu _1= 1.02$, $\nu _2=0.96 $, $\nu _3=1.02 $,
$\delta _1=0 $, $\delta _2=3.142 $, $\delta _3=0.0420 $; b) $\nu _1=
1.02$, $\nu _2=0.96 $, $\nu _3=1.02 $, $\delta _1=0 $, $\delta _2=0.0
$, $\delta _3=0.0420 $.\label{fig:III(ii)-(vii)}}

\end{figure}

In the last two figures \ref{fig:III(ii)-(vii)}  we used W2
set of soliton widths and a choice of parameters characteristic for case
III, i.e. $p $ is real while $q $ is purely imaginary. In
Fig.~\ref{fig:III(ii)-(vii)}a $Q>0 $ with  BSR, and in
Fig.~\ref{fig:III(ii)-(vii)}b, we have $Q<0 $ and  AFR.

\section{Conclusions}\label{sec:7}

In this paper we extend the formalism by Karpman and Solov'ev
proposed to describe the NSE 2-soliton interaction~\cite{KS} and
generalized to arbitrary number of solitons
~\cite{Gerd1,Gerd2,Gerd3,Gerd4}, to the case of the modified
nonlinear Schr\"odinger equation. The aim of our paper was
two-fold. First, we would like to investigate a possibility to
apply an integrable chain-like model to capture adiabatic
dynamics of MNSE solitons within the $N$-soliton train. Because a
functional form of the MNSE soliton is not of the familiar
hyperbolic-secant type with a real argument, we might expect an
existence of some novel features as compared with the NSE case.
We show that, under specific well-defined conditions, the
dynamical system of $4N$ equations for soliton parameters is
reduced to the completely integrable complex Toda chain model
with $N$ nodes. This is a strong argument in favor of
universality of the CTC model for $N$-soliton interactions. Though
the same CTC arises also for the NSE, there are a few
peculiarities inherent in the MNSE solitons. In particular, we
found out more complicated phase behavior of the $N$-soliton
train. Using the integrability of the CTC, we are able to predict
various asymptotic regimes of the MNSE $N$-soliton train
evolution. Besides, we point out the sets of the initial soliton
parameters corresponding to each of the dynamical regimes.
Numerical simulations of the MNSE 2- and 3-soliton interactions
are in very good agreement with the CTC-based predictions.
Evidently, the results obtained can be extended to treat also
multicomponent (vector) generalizations of both the NSE (see,
e.g., \cite{Yang,Yang2,VS-36} and references therein) and MNSE
~\cite{Hisakado1,Hisakado2,Dok-Jpn}. Work in this direction is
now in progress. We note that in non-integrable wave systems,
Toda-chain type equations may still be
derived for the adiabatic interaction of $N$ nearly identical
solitary waves, but such equations are generally non-integrable
as well \cite{Arn,Yang2}.

Secondly, we consider the MNSE as a true starting integrable
model to describe subpicosecond pulse evolution in nonlinear
media. Strictly speaking, to justify a relevance of our results
to actual ultrashort pulses, we should also account in our model
at least two additional effects, the third-order dispersion and
intrapulse Raman scattering. These effects break the
integrability of the CTC, and we are faced with a truly perturbed
MNSE. Following the lines of recently established interrelations
between the perturbed NSE and perturbed CTC~\cite{Gerd4}, we can
extend the above formalism to account for small actual
perturbations which act along with the effective perturbation
(\ref{eq:14}). The corresponding results will be published
elsewhere. The single MNSE soliton dynamics in the presence of
the intrapulse Raman scattering is discussed in the recent
paper~\cite{Afan}.

Recently we were informed~\cite{Val} that the CTC model arises
also in the case of the soliton-train propagation in a system
governed by the classical Thirring model~\cite{Thir,Kuzn}.  This
seems natural in view of the facts that: i)~CTC describes the
adiabatic soliton interactions for all NLEE of the NLS hierarchy;
ii)~MTM is just another representative of the MNLS hierarchy.

There remain several natural questions that will be addressed in
sequels of this paper. The first one is the limit $\alpha \to 0 $
in which we should recover the results for the NSE $N $-soliton
trains. We have proved that the Karpman-Solov'ev-like equations
for MNSE $N$ solitons transform under this limit to the known NSE
formulae. The second one concerns the treatment of the perturbed
versions of the MNSE and the corresponding perturbed CTC model;
for the NSE such perturbed CTC models have been briefly analyzed
in \cite{Gerd4}.

\section*{Acknowledgements}\label{sec:Ack}

The authors gratefully acknowledge many stimulating discussions
with Dr. V. Shchesnovich. Substantial part of this work was done
during the  RCP 264 Conference held in June 2000 in Montpellier.
V.G. and E.D. are grateful to Professors J.-G. Caputo and P.
Sabatier for their support which made our participation in this
conference possible. The work of E.D. was supported in part by
the Grant F98-044 from the Belarussian Foundation for Fundamental
Research. The work of J.Y.
was supported in part by the Air Force Office of Scientific
Research under contract F49620-99-1-0174,
and by the National Science Foundation under grant DMS-9971712.

\newpage
\listoffigures

\end{document}